\pdfoutput=1

\documentclass[11pt,twoside,a4paper,cmspaper,final,collab]{cms-tdr}
\def\svnVersion{363173}\def\cmsCernNoTag{CERN-PH-EP/2015-331}\def\cmsCernDate{\today}\def\cmsMessage{Published in Physics Letters B as \href{http://dx.doi.org/10.1016/j.physletb.2016.06.004}{\doi{10.1016/j.physletb.2016.06.004.}}}

\begin{document}\cmsNoteHeader{HIG-14-035}

\hyphenation{had-ron-i-za-tion}
\hyphenation{cal-or-i-me-ter}
\hyphenation{de-vices}
\RCS$Revision: 362673 $
\RCS$HeadURL: svn+ssh://svn.cern.ch/reps/tdr2/papers/HIG-14-035/trunk/HIG-14-035.tex $
\RCS$Id: HIG-14-035.tex 362673 2016-08-02 20:22:20Z jstupak $
\newlength\cmsFigWidth
\ifthenelse{\boolean{cms@external}}{\setlength\cmsFigWidth{0.85\columnwidth}}{\setlength\cmsFigWidth{0.4\textwidth}}
\ifthenelse{\boolean{cms@external}}{\providecommand{\cmsLeft}{top\xspace}}{\providecommand{\cmsLeft}{left\xspace}}
\ifthenelse{\boolean{cms@external}}{\providecommand{\cmsRight}{bottom\xspace}}{\providecommand{\cmsRight}{right\xspace}}

\newcommand{\PV}{\ensuremath{\mathrm{V}}\xspace}
\newcommand{\Pj}{\ensuremath{\mathrm{j}}\xspace}

\newcommand{\zerop}{\ensuremath{0^+}\xspace}
\newcommand{\zerom}{\ensuremath{0^-}\xspace}
\newcommand{\HtoVV}{\ensuremath{\ensuremath{\PH\to\PV\PV}}\xspace}
\newcommand{\HtoWW}{\ensuremath{\ensuremath{\PH\to\PW\PW}}\xspace}
\newcommand{\HtoZZ}{\ensuremath{\ensuremath{\PH\to\PZ\PZ}}\xspace}
\newcommand{\ucm}{uncorrelated-$\mu$\xspace}
\renewcommand{\cm}{correlated-$\mu$\xspace}
\newcommand{\faThree}{\ensuremath{f_{a_3}}\xspace}
\newcommand{\faThreeZH}{\ensuremath{\faThree^{\PZ\PH}}\xspace}
\newcommand{\faThreeZZ}{\ensuremath{\faThree^{\PZ\PZ}}\xspace}
\newcommand{\faThreeWH}{\ensuremath{\faThree^{\PW\PH}}\xspace}
\newcommand{\faThreeWW}{\ensuremath{\faThree^{\PW\PW}}\xspace}
\newcommand{\faThreeVV}{\ensuremath{\faThree^{\PV\PV}}\xspace}
\newcommand{\Wh}{\ensuremath{\PW\PH}\xspace}
\newcommand{\Zh}{\ensuremath{\PZ\PH}\xspace}
\newcommand{\Vh}{\ensuremath{\PV\PH}\xspace}
\newcommand{\WH}{\Wh}
\newcommand{\ZH}{\Zh}
\newcommand{\Wenu}{\ensuremath{\PW\to\Pe\nu}\xspace}
\newcommand{\Wmunu}{\ensuremath{\PW\to\mu\nu}\xspace}
\newcommand{\Wlnu}{\ensuremath{\PW\to\ell\nu}\xspace}
\newcommand{\Zee}{\ensuremath{\PZ\to\Pe\Pe}\xspace}
\newcommand{\Zmumu}{\ensuremath{\PZ\to\mu\mu}\xspace}
\newcommand{\Zll}{\ensuremath{\PZ\to\ell\ell}\xspace}
\newcommand{\JHUGEN}{\textsc{JHUGen}\xspace}
\newcommand{\mVh}{\ensuremath{m\!\left(\PV\PH\right)}\xspace}
\newcommand{\mVH}{\mVh} 

\cmsNoteHeader{HIG-14-035}
\title{Combined search for anomalous pseudoscalar $\PH\PV\PV$ couplings in $\PV\PH(\PH\to\PQb\PAQb)$ production and $\PH\to\PV\PV$ decay}

\date{\today}

\abstract{
A search for anomalous pseudoscalar couplings of the Higgs boson $\PH$ to electroweak vector bosons $\PV$ ($=\PW$ or $\PZ$) in a sample of proton-proton collision events corresponding to an integrated luminosity of 18.9\fbinv at a center-of-mass energy of 8\TeV is presented.  Events consistent with the topology of associated $\PV\PH$ production, where the Higgs boson decays to a pair of bottom quarks  and the vector boson decays leptonically, are analyzed. The consistency of data with a potential pseudoscalar contribution to the $\PH\PV\PV$ interaction, expressed by the effective pseudoscalar cross section fractions \faThree, is assessed by means of profile likelihood scans. Results are given for the $\PV\PH$ channels alone and for a combined analysis of the $\PV\PH$ and previously published \HtoVV channels.  Under certain assumptions,
$\faThreeZZ>0.0034$ is excluded at 95\% confidence level in the combination.  Scenarios in which these assumptions are relaxed are also considered.
}

\hypersetup{%
pdfauthor={CMS Collaboration},%
pdftitle={Combined search for anomalous pseudoscalar HVV couplings in VH (H to bb) production and H to VV decay},%
pdfsubject={CMS},%
pdfkeywords={CMS, physics, Higgs, BSM}}

\maketitle

\section{Introduction}

The observation of a new boson~\cite{Aad:2012tfa,Chatrchyan:2012ufa,Chatrchyan:2013lba} with a mass around $125\GeV$ and properties consistent with those of the standard model (SM) Higgs boson~\cite{StandardModel67_1,Englert:1964et,Higgs:1964ia,Higgs:1964pj,Guralnik:1964eu,StandardModel67_2,StandardModel67_3} has ushered in a new era of precision Higgs physics.  The ATLAS and CMS collaborations at the CERN LHC have begun a comprehensive study of the boson properties.
The spin-parity of the Higgs boson has been studied in  $\PH\to\PZ\PZ$, $\PZ\gamma^{*}$,$\gamma^{*}\gamma^{*}\to4\ell$, $\PH\to\PW\PW\to \ell\nu\ell\nu$, and $\PH\to\gamma\gamma$ decays~\cite{Chatrchyan:2012jja,Chatrchyan:2013mxa,Chatrchyan:2013iaa,Khachatryan:2014ira,Khachatryan:2014kca,Aad:2013xqa}, where $\ell$ is an electron or muon.
The CDF and D0 collaborations have set limits on the $\Pp \PAp \to\PV\PH$ production cross section (with $\PV=\PW$ or $\PZ$) at the Tevatron, for two exotic spin-parity models of the Higgs boson~\cite{Aaltonen:2015mka}.
In all cases, the spin-parity $J^{CP}$ of the boson has been found to be consistent with the SM prediction.  Based on a study of anomalous couplings in $\PH\to\PZ\PZ\to 4\ell$ decays, the CMS collaboration has excluded the hypothesis of a pure pseudoscalar spin-zero boson at 99.98\% confidence level (CL), while an effective pseudoscalar cross section fraction $\faThreeZZ>0.43$ is excluded at 95\% CL (assuming a positive, real valued ratio of scalar and pseudoscalar couplings)~\cite{Khachatryan:2014kca}.  Under the same assumptions, the ATLAS collaboration has excluded $\faThreeZZ>0.11$ at 95\% CL~\cite{Aad:2015mxa}.

We present here the first search for anomalous pseudoscalar HVV couplings at the LHC in the topology of associated production, VH.
It will be shown that the VH channels are strong probes of the structure of the HVV interaction, with sensitivity even to small anomalous couplings.  The ultimate LHC sensitivity to a potential pseudoscalar interaction in these channels is expected to greatly exceed that of $\PH\to\PV\PV$~\cite{Anderson:2013afp}.
Due to the highly off-shell nature of the propagator in VH production, small anomalous couplings can lead to significant modifications of cross sections and kinematic features.
In particular, the propagator mass, measured by the VH invariant mass, \mVH, is highly sensitive to anomalous $\PH\PV\PV$ couplings~\cite{Ellis:2012xd}.

Results from the VH channels are ultimately combined with those from \HtoVV measurements~\cite{Khachatryan:2014kca}.
The $\PQq\PAQq\to\PV\PH\to\PV\PQb\PAQb$ and $\Pg\Pg\to\PH\to\PV\PV$ processes involve the Yukawa fermion coupling $\PH\mathrm{f\overline{f}}$ and the same $\PH\PV\PV$ coupling, assuming gluon fusion production is dominated by the top-quark loop. The dominance of the gluon fusion production mechanism of the Higgs boson at the LHC is supported by experimental measurements~\cite{StandardModel67_1,Englert:1964et,Higgs:1964ia,Higgs:1964pj,Guralnik:1964eu,StandardModel67_2,StandardModel67_3}. It is interesting to consider models where the ratio of the $\PH\PQb\PAQb$ and $\PH\ttbar$ coupling strengths in the VH and \HtoVV processes is not affected by the presence of anomalous contributions~\cite{Barnett1984191}. In such a case, it is possible to relate the cross sections of the two processes for arbitrary anomalous $\PH\PV\PV$ couplings and perform a combined analysis of the VH and \HtoVV processes, exploiting both kinematics and the relative signal strengths of the two processes.  The \HtoVV signal strength is relatively well measured and can provide a strong constraint on the VH signal strength.  For modest values of \faThreeZZ, the VH signal strength is constrained to large values.  The added constraint thereby significantly improves the sensitivity to anomalous couplings.

In the following, we consider only the interactions of a spin-zero boson with the W and Z bosons, for which the scattering amplitude is parameterized as
\begin{equation}
\ifthenelse{\boolean{cms@external}}
{
\begin{split}
 A(\PH\PV\PV) \sim & \left[ a_{1}^{\PH\PV\PV}
  + \frac{\kappa_1^{\PH\PV\PV}q_{\PV_1}^2  + \kappa_2^{\PH\PV\PV} q_{\PV_2}^{2}}{\left(\Lambda_{1}^{\PH\PV\PV} \right)^{2}} \right]
 m_{\PV_1}^2 \epsilon_{\PV_1}^* \epsilon_{\PV_2}^*  \\
 & + a_{2}^{\PH\PV\PV}  f_{\mu \nu}^{*(1)}f^{*(2)\mu\nu}
 + a_{3}^{\PH\PV\PV}   f^{*(1)}_{\mu \nu} {\tilde f}^{*(2)\mu\nu},
\end{split}
}
{
 A(\PH\PV\PV) \sim
 \left[ a_{1}^{\PH\PV\PV}
 + \frac{\kappa_1^{\PH\PV\PV}q_{\PV_1}^2 + \kappa_2^{\PH\PV\PV} q_{\PV_2}^{2}}{\left(\Lambda_{1}^{\PH\PV\PV} \right)^{2}} \right]
 m_{\PV_1}^2 \epsilon_{\PV_1}^* \epsilon_{\PV_2}^*
 + a_{2}^{\PH\PV\PV}  f_{\mu \nu}^{*(1)}f^{*(2)\mu\nu}
 + a_{3}^{\PH\PV\PV}   f^{*(1)}_{\mu \nu} {\tilde f}^{*(2)\mu\nu},
}
 \label{eq:formfact-fullampl-spin0}
\end{equation}
where the $a_{i}^{\PH\PV\PV}$ are arbitrary complex coupling parameters which can depend on the V$_1$  and V$_2$ squared four-momenta, $q_{\PV_1}^2$ and $q_{\PV_2}^2$; $f^{(i){\mu \nu}}$ is the field strength tensor of a gauge boson with momentum $q_{\PV_i}$ and polarization vector $\epsilon_{\PV_i}$, given by $\epsilon_{\PV_i}^{\mu}q_{\PV_i}^{\nu} - \epsilon_{\PV_i}^\nu q_{\PV_i}^{\mu} $; ${\tilde f}^{(i)}_{\mu \nu}$ is the dual field strength tensor, given by $\frac{1}{2} \epsilon_{\mu\nu\rho\sigma} f^{(i)\rho\sigma}$; $m_{\PV_1}$ is the pole mass of the vector boson; and $\Lambda_{1}^{\PH\PV\PV}$ is the energy scale where phenomena not included in the SM become relevant~\cite{Anderson:2013afp}.  The $a_1^{\PH\PV\PV}$, $\kappa_i^{\PH\PV\PV}$ and $a_2^{\PH\PV\PV}$ terms represent parity-conserving interactions of a scalar, while the $a_3^{\PH\PV\PV}$ term represents a parity-conserving interaction of a pseudoscalar.  In the SM, $a_1^{\PH\PV\PV}=2$, which is the only nonzero coupling at tree level.  All other terms in Eq.~(\ref{eq:formfact-fullampl-spin0}) are generated within the SM by loop-induced processes at levels below current experimental sensitivity.  Therefore, any evidence for these terms in the available data should be interpreted as evidence of new physics.

We search for an anomalous $a_3^{\PH\PV\PV}$ term of the $\PH\PV\PV$ interaction, assuming that the $\kappa_i^{\PH\PV\PV}$ and $a_2^{\PH\PV\PV}$ terms are negligible.  Throughout the remainder of the paper, the term ``scalar interaction`` will be used to describe the $a_1^{\PH\PV\PV}$ term.
The effective pseudoscalar cross section fraction for process $j$ (WH, ZH, WW, or ZZ) is defined as
\begin{equation}
\faThree^j = \frac{{\left|a_3^{\PH\PV\PV}\right|}^2 \sigma^j_{3}}{{\left|a_1^{\PH\PV\PV}\right|}^2 \sigma^j_{1} + {\left|a_3^{\PH\PV\PV}\right|}^2 \sigma^j_{3}},
\label{eq:fa_definitions}
\end{equation}
where $\sigma_i^j$ is the production cross-section for process $j$ with $a_i^{\PH\PV\PV}=1$ and all other couplings assumed to be equal to zero.  A superscript is not included when making a general statement not related to a particular process.  The purely scalar (pseudoscalar) case corresponds to $\faThree=0$ ($\faThree=1$).  The signal strength parameter $\mu^j$ for process $j$ can also be defined in terms of the $a_i^{\PH\PV\PV}$ as
\begin{equation}
\mu^j=\frac{{\left|a_1^{\PH\PV\PV}\right|}^2 \sigma^j_{1} + {\left|a_3^{\PH\PV\PV}\right|}^2 \sigma^j_{3}}{{\left|a_{1,\mathrm{SM}}^{\PH\PV\PV}\right|}^2\sigma^j_{1}}.
\end{equation}
For a given set of coupling constants, the physical observables $\faThree^j$ and $\mu^j$ vary for different processes as a result of the dependence on the $\sigma_i^j$.
The \faThreeZH and \faThreeWH variables are defined with respect to the ZH and WH production cross-sections in $\sqrt{s}=8~\TeV$ pp collisions, whereas the \faThreeVV variables are defined with respect to the cross-section times branching fraction for the corresponding $\Pp\Pp\to\HtoVV$ process.  In the latter case, the dependence on the $\Pp\Pp\to\PH$ cross-section cancels.

\section{The CMS detector}
\label{sec:CMS}

The central feature of the CMS apparatus is a superconducting solenoid of 6\unit{m} internal diameter, providing a magnetic field of 3.8\unit{T}. Within the solenoid volume are a silicon pixel and strip tracker, a lead tungstate crystal electromagnetic calorimeter, and a brass and scintillator hadron calorimeter, each composed of a barrel and two endcap sections. Extensive forward calorimetry complements the coverage provided by the barrel and endcap detectors.  Muons are measured in gas-ionization detectors embedded in the steel flux-return yoke outside the solenoid. A more detailed description of the CMS detector, together with a definition of the coordinate system used and the relevant kinematic variables, can be found in Ref.~\cite{Chatrchyan:2008zzk}.

\section{Analysis strategy}
\label{sec:strat}

The analysis is based on a data sample of pp collisions corresponding to an integrated luminosity of 18.9\fbinv at a center-of-mass energy of 8\TeV, collected with single-electron, single-muon, and double-electron triggers.  The final states considered are $\ell\nu\Pj\Pj$ and $\ell\ell \Pj\Pj$ (where $\Pj$ represents a jet), targeting the WH and ZH signals respectively.

The trigger, object and event selection criteria, and background modeling are identical to those of Ref.~\cite{Chatrchyan:2013zna}.  Using the selected events, the two-dimensional template method described in Ref.~\cite{Khachatryan:2014kca} is used to determine \faThree confidence intervals.
The discriminant of the boosted decision tree (BDT) described in Ref.~\cite{Chatrchyan:2013zna} serves as one dimension of the templates.  This BDT is trained separately for the WH and ZH channels to exploit various kinematic features typical of signal and background, and the correlations among observables.  The b-tagging likelihood discriminants of the jets used to construct the Higgs boson candidate, the invariant mass of the Higgs boson candidate, and the angular separation between final state leptons and jets are the most important variables in terms of background rejection.  Although initially trained to separate background from a scalar Higgs boson signal, it has been demonstrated with simulated events that the BDT is also effective for signals with anomalous \faThree values.
The second dimension of the templates is \mVh.  Effectively, the BDT dimension provides a background-depleted region at high values of the BDT discriminant with which to test various signal hypotheses using the \mVh distribution.

Signal templates in the $\vec{x} = \left\{\mathrm{BDT},\mVh \right\}$ plane are constructed for arbitrary values of \faThree from a linear superposition of templates representing the pure scalar ($\mathcal{P}_{0^+}\left(\vec{x}\right)$) and pseudoscalar ($\mathcal{P}_{0^-}\left(\vec{x}\right)$) hypotheses and a template ($\mathcal{P}^\text{int}_{0^+,0^-}\left(\vec{x}; \phi_{a_3}\right)$) that accounts for interference between the $a_1^{\PH\PV\PV}$ and $a_3^{\PH\PV\PV}$ terms in Eq.~(\ref{eq:formfact-fullampl-spin0}), as follows:
\begin{equation}
\ifthenelse{\boolean{cms@external}}
{
\begin{split}
\mathcal{P}_\text{sig}\left(\vec{x}; \faThree,\phi_{a_3}\right)  = &  \left(1-\faThree\right) \, \mathcal{P}_{0^+}\left(\vec{x}\right)
 + \faThree \, \mathcal{P}_{0^-}\left(\vec{x}\right) \\
 & +   \sqrt{\faThree\left(1-\faThree\right)}\, \mathcal{P}^\text{int}_{0^+,0^-}\left(\vec{x}; \phi_{a_3}\right).
\end{split}
}
{
\mathcal{P}_\text{sig}\left(\vec{x}; \faThree,\phi_{a_3}\right) =  \left(1-\faThree\right) \, \mathcal{P}_{0^+}\left(\vec{x}\right)
+ \faThree \, \mathcal{P}_{0^-}\left(\vec{x}\right)
+ \sqrt{\faThree\left(1-\faThree\right)}\, \mathcal{P}^\text{int}_{0^+,0^-}\left(\vec{x}; \phi_{a_3}\right).
}
\label{eq:fractions-general}
\end{equation}
The phase between the $a_1^{\PH\PV\PV}$ and $a_3^{\PH\PV\PV}$ couplings is represented by $\phi_{a_3}$.  The interference contributions to the BDT discriminant and \mVh distributions are negligible, as verified with simulated events.
Therefore the last term in Eq.~(\ref{eq:fractions-general}) is ignored in the VH channels.
Equation~(\ref{eq:fractions-general}) is also used to parameterize the \HtoVV signals.
Anomalous couplings that result from loops with particles much heavier than the Higgs boson are real valued, allowing phases of 0 and $\pi$.  In the \HtoVV channels, we assume $\phi_{a_3}=0$.
The resulting templates are used to perform profile likelihood scans~\cite{ATL-PHYS-PUB-2011-011} to assess the consistency of various signal hypotheses with the data.  One-dimensional profile likelihood scans of \faThree are performed (where $\mu$ is profiled), as well as two-dimensional scans in the $\mu$ versus \faThree plane.

In order to combine channels that depend on the $a_i^{\PH\PZ\PZ}$ with those depending on the $a_i^{\PH\PW\PW}$, some assumption on the relationship between the couplings is required, and custodial symmetry is assumed ($a_1^{\PH\PZ\PZ}=a_1^{\PH\PW\PW}$).  It is further assumed that $a_3^{\PH\PW\PW}=a_3^{\PH\PZ\PZ}$.
With these assumptions, the \faThree and $\mu$ values in the \WH and \ZH channels are related by
\begin{equation}
\faThreeWH = \left[ 1+\frac{1}{\Omega^{\PZ\PH,\PW\PH}}\left(\frac{1}{\faThreeZH}-1\right)\right]^{-1}
\label{eq:faThreeConversion}
\end{equation}
and
\begin{equation}
\mu^{\PW\PH}=\mu^{\PZ\PH}\, \left[1+\faThreeZH\left( \Omega^{\PZ\PH,\PW\PH}-1\right)\right],
\label{eq:muConversion}
\end{equation}
where
\begin{equation}
\Omega^{\PZ\PH,\PW\PH}=\frac{\sigma_1^{\PZ\PH}/\sigma_3^{\PZ\PH}}{\sigma_1^{\PW\PH}/\sigma_3^{\PW\PH}}.
\label{eq:omega}
\end{equation}
The $\sigma_1/\sigma_3$ ratios given by the \JHUGEN 4.3~\cite{Anderson:2013afp,Bolognesi:2012mm,Gao:2010qx} event generator and values of $\Omega^{i,j}$ are given in Tables~\ref{tab:translationConst} and \ref{tab:Omega}, respectively.
\begin{table}[tbp]
\renewcommand{\arraystretch}{1.2}
  \topcaption{$\sigma_1/\sigma_3$ cross section ratios calculated with \JHUGEN.}
  \label{tab:translationConst}
  \centering
    \newcolumntype{,}{D{,}{\,,\,}{-1}}
    \begin{tabular}{,c} \hline
        \multicolumn{1}{c}{Process}  & $\sigma_1/\sigma_3$ \\  \hline
        \PW\PH  & 0.0174 \\
        \PZ\PH  & 0.0239 \\
        \PW\PW  & 3.01 \\
        \PZ\PZ & 6.36 \\
    \hline
    \end{tabular}
\end{table}
\begin{table}[tbp]
\renewcommand{\arraystretch}{1.2}
  \topcaption{Values of $\Omega^{i,j}$ which relate the channels studied in this paper, as defined in Eq.~(\ref{eq:omega}).}
  \label{tab:Omega}
  \centering
    \newcolumntype{,}{D{,}{\,,\,}{-1}}
    \begin{tabular}{,c} \hline
	i,j 					& $\Omega^{i,j}$ \\  \hline
	\PZ\PH, \PW\PH 	& 1.37 \\
	\PZ\PZ, \PW\PW 	& 2.11 \\
	\PZ\PZ, \PZ\PH 	& 266 \\
	\PW\PW, \PW\PH & 173 \\
    \hline
    \end{tabular}
\end{table}
In order to improve the sensitivity to anomalous couplings, results from the VH channels are combined with those from \HtoVV~\cite{Khachatryan:2014kca}.  We assume the signal yield in the \HtoVV analysis to be dominated by gluon fusion production with negligible contamination from vector boson fusion or VH production, as in Ref.~\cite{Khachatryan:2014kca}.  Provided that the ratio of the $\PH \PQb \PAQb$ and $\PH \PQt \PAQt$ coupling strengths is given by the SM prediction, Eq.~(\ref{eq:muConversion}) can be used to relate the signal strength in the VH and \HtoVV analyses, with an appropriate change of indices (replacing `WH` with `ZZ` to relate the $\PZ\PZ$ and $\PZ\PH$ channels, or `ZH` with `WW` to relate the $\PW\PW$ and $\PW\PH$ channels).
In the combination of the WH and \HtoWW channels, the ratio of the signal strengths $\mu^{\PW\PH}/\mu^{\PW\PW}$ increases linearly from 1 to 173 as \faThreeWW increases from 0 to 1, according to Eq.~(\ref{eq:muConversion}).  The WH signal strength has been measured by CMS to be $1.1\pm 0.9$~\cite{Chatrchyan:2013zna}, and for \HtoWW it has been measured to be $0.76\pm 0.21$~\cite{Chatrchyan:2013iaa}.  Thus, for intermediate and large values of \faThreeWW it is not possible to reconcile the expected signal yield with data in both channels simultaneously.  A similar effect occurs in a combination of the ZH and \HtoZZ channels, where the ratio of the signal strengths $\mu^{\PZ\PH}/\mu^{\PZ\PZ}$ rises sharply with \faThreeZZ.

However, an anomalous ratio of the $\PH\PQb\PAQb$ and $\PH\PQt\PAQt$ coupling strengths spoils the relationship in Eq.~(\ref{eq:muConversion}).  We therefore perform two interpretations of the VH and \HtoVV combination; one interpretation in which this relationship is enforced, and one interpretation in which the signal strengths in the $\PV\PH$ and \HtoVV channels are allowed to vary independently.  These are referred to as the `\cm' and `\ucm' combinations, respectively.

\section{Simulation}
\label{sec:MC}

Simulated $\PQq\PQq\to\PV\PH$ signal events are generated for pure scalar and pseudoscalar hypotheses with the leading-order (LO) event generator \JHUGEN, and assuming a mass $m_{\PH}=125.6\GeV$.
The simulated event sample is reweighted based on the vector boson $p_T$ to include corrections up to next-to-next-to-LO and next-to-LO (NLO) in the QCD and electroweak (EW) couplings respectively~\cite{Han:1991ia,vanNeerven199211,Brein2012,Dittmaier:2012vm,Ciccolini:2003jy}.  These corrections are derived for a scalar Higgs boson, and applied to both scalar and pseudoscalar simulated event samples.

\begin{figure}[htbp]
  \centering
  \ifthenelse{\boolean{cms@external}}
  {
  \includegraphics[width=0.47\textwidth,keepaspectratio]{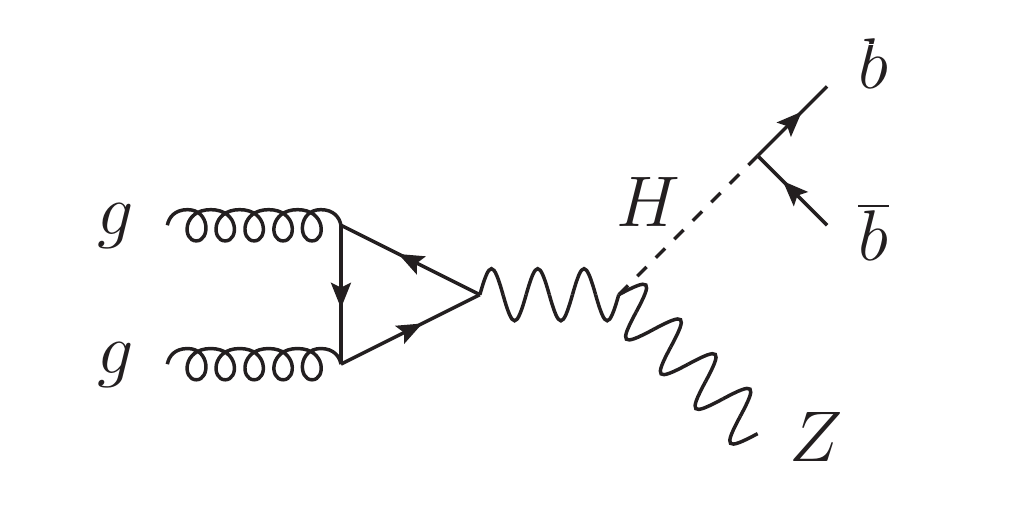}
  \includegraphics[width=0.47\textwidth,keepaspectratio]{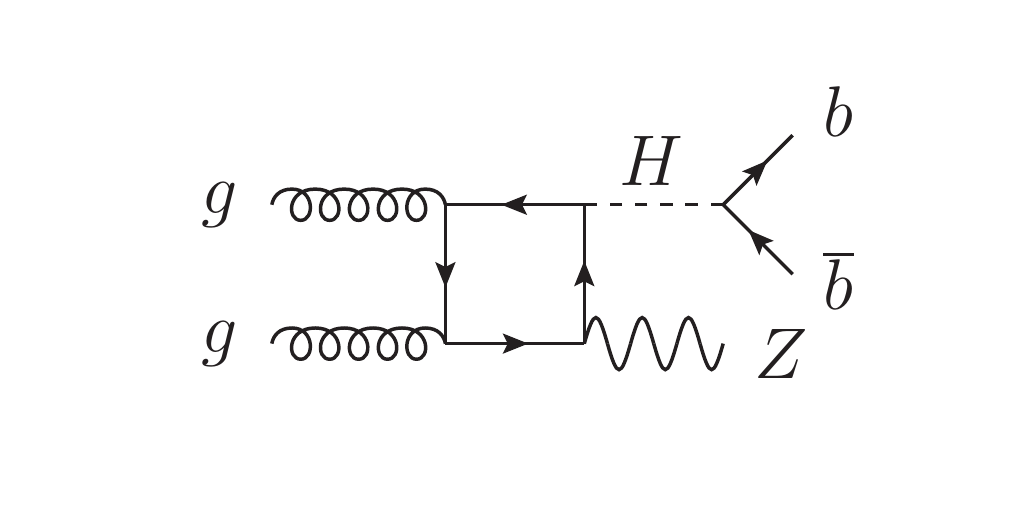}
  }
  {
        \begin{tabular}{cc}
                \raisebox{-.5\height}{\includegraphics[width=0.47\textwidth,keepaspectratio]{Figure_001-a.pdf}} &
                \raisebox{-.5\height}{\includegraphics[width=0.47\textwidth,keepaspectratio]{Figure_001-b.pdf}} \\
                \end{tabular}
  }
  \caption{Feynman diagrams representing gluon-initiated ZH production via a quark triangle (\cmsLeft) and box (\cmsRight) loop.}
  \label{fig:feynman}
\end{figure}

The $\Pg\Pg\to\PZ\PH$ process includes diagrams with quark triangle and box loops, as shown in Fig.~\ref{fig:feynman}.  These diagrams interfere destructively with one another~\cite{Englert:2013vua}.  The box diagram contains no $\PH\PV\PV$ vertex.  The triangle diagram does, but is unaffected by the $a_3^{\PH\PV\PV}$ term in Eq.~(\ref{eq:formfact-fullampl-spin0}).  The triangle diagram mediated by a CP-odd $\PH\PV\PV$ interaction is completely anti-symmetric under the reversal of the direction of loop momentum flow; the diagrams with opposite loop momentum flow therefore perfectly cancel one another.  As the $a_1^{\PH\PZ\PZ}$ coupling varies within a profile likelihood scan, the box contribution remains fixed while the triangle contribution and the interference must be varied accordingly.  This is accomplished by reweighting the simulated $\Pg\Pg\to\PZ\PH$ event sample to have the correct \mVH distribution at the generator level, including interference effects.  This reweighting is based on results obtained with the VBFNLO event generator~\cite{Englert:2013vua,Arnold:2008rz}, modified for this analysis to allow variation of the $\PH\mathrm{f\overline{f}}$ and $\PH\PZ\PZ$ coupling strengths.

Simulated background event samples are generated with a variety of event generators.
Diboson, W+jets, Z+jets, and \ttbar samples are generated with \MADGRAPH~5.1~\cite{madgraph}, while \POWHEG~1.0~\cite{powheg} is used to generate single top quark samples, as well as the gluon-initiated contribution to ZH production ($\Pg\Pg\to\PZ\PH$).  The \HERWIGpp~2.5~\cite{herwig} generator is used along with alternative matrix element generators to produce additional simulated background samples to assess the systematic uncertainty related to event simulation accuracy, as described in Section~\ref{sec:systematics}.

The \PYTHIA 6.4~\cite{pythia} and \HERWIGpp generators are used to simulate parton showering and hadronization.  Detector simulation is performed with \GEANTfour~\cite{Agostinelli2003250}.  Uncorrelated proton-proton collisions occurring in the same bunch crossing as the signal event (pileup) are overlayed on top of the hard interaction, in accord with the distribution observed.  Corrections are applied to the simulation in order to account for differences in object reconstruction efficiencies and resolutions with respect to the data.

Control regions in data are defined in Ref.~\cite{Chatrchyan:2013zna}, from which normalization scale factors for the dominant backgrounds are derived.  A simultaneous fit to data across control regions is performed to extract the scale factors, which are applied here.
The shape of the \PW (\PV) boson transverse momentum \pt distribution is corrected in the simulated \ttbar (V+jets) event sample, based on a fit to data in a background-enriched control region.

\section{Object and event selection}
\label{sec:selection}

All objects are reconstructed using a particle-flow (PF) approach~\cite{CMS-PAS-PFT-09-001, CMS-PAS-PFT-10-001}.  Among all reconstructed primary vertices satisfying basic quality criteria, the vertex with the largest value of $\sum \pt^2$ is selected.
Electrons are reconstructed from inner detector tracks matched to calorimeter superclusters, and selected with a multivariate identification algorithm~\cite{Khachatryan:2015hwa}.  Electrons are required to have $\pt>30\GeV$ and pseudorapidity $\abs{\eta}<2.5$, with a veto applied to the barrel-endcap transition region ($1.44<\abs{\eta}<1.57$) where electron reconstruction is sub-optimal.  Muons are reconstructed from inner detector tracks matched to tracks reconstructed in the muon system, and selected with a cut-based identification algorithm~\cite{Chatrchyan:2012xi}.  Muons are required to have $\pt>20\GeV$ and $\abs{\eta}<2.4$.  Both electrons and muons are required to be well isolated from other reconstructed objects.  Jets are reconstructed using the anti-\kt algorithm~\cite{Cacciari:2008gp}, with a distance parameter of 0.5, from the reconstructed objects, after removing charged objects with a trajectory inconsistent with production at the primary vertex.  Additionally, the energy contribution from neutral pileup activity is subtracted with an area-based approach~\cite{Cacciari:2007fd}.  Jets are tagged as originating from the fragmentation and hadronization of bottom quarks with the combined secondary vertex (CSV) algorithm~\cite{Chatrchyan:2012jua}, which exploits both the track impact parameter and secondary vertex information.  Missing transverse energy $\MET$ is reconstructed as the negative vector \pt sum of all reconstructed objects.

Events are categorized based on the flavour and number of charged leptons into four channels.  Events with two same-flavour, opposite-sign electrons (muons) are assigned to the \Zee (\Zmumu) channel.  Events with one electron (muon) and large \MET are assigned to the \Wenu (\Wmunu) channel.  In the \Wlnu (\Zll) channels, Higgs boson candidates are constructed from the pair of jets (referred to as $\Pj_1$ and $\Pj_2$) with the largest vector \pt sum among jets with $\pt>30$ (20)\GeV and $\abs{\eta}<2.5$. The Z boson candidates are constructed from lepton pairs whose invariant mass is consistent with the Z boson mass.  The W boson candidates are constructed by combining the momentum of the identified lepton with the event \MET, and calculating the neutrino momentum along the beam axis based on a W boson mass constraint.  To suppress contributions from QCD multijet events, in the \Wlnu channels the magnitude of the \MET vector must exceed $45\GeV$ and it must be separated in direction from the charged lepton by less than $\pi/2$~radians in azimuth.  In addition, the Higgs boson candidate \pt must exceed $100\GeV$.

The analysis sensitivity is increased further by categorizing events into medium- and high-boost regions based on the \pt of the vector boson candidate.  The bulk of the sensitivity comes from the high-boost region.  These regions are later combined statistically.
In the \Wlnu channels, the medium- and high-boost regions are defined by $130<\pt(\PW)<180\GeV$ and $\pt(\PW)>180\GeV$, respectively.
In the \Zll channels, the regions are instead defined by $50<\pt(\PZ)<100\GeV$ and $\pt(\PZ)>100\GeV$.
The low-boost region described in Ref.~\cite{Chatrchyan:2013zna} is not included because of its negligible sensitivity to anomalous couplings.  Requirements on the Higgs boson candidate mass and the b-tagging likelihood discriminants of the jets used to construct the Higgs boson candidate are also applied.  The selection criteria are summarized in Table~\ref{tab:BDTsel}.

\begin{table*}[tbp]
  \topcaption{Summary of the event selection criteria.  Numbers in parentheses refer to the high-boost region defined in the text.}
  \label{tab:BDTsel}
  \centering
  \renewcommand{\arraystretch}{1.2}
    \newcolumntype{x}{D{[}{\, [}{-1}}
    \begin{tabular}{xcc} \hline
      \multicolumn{1}{c}{Variable}                     & \Wlnu                             & \Zll                  \\ \hline
      \pt(\Pj_1) [\GeVns]                   & ${>}30$                           & ${>}20$                  \\
      \pt(\Pj_2) [\GeVns]                   & ${>}30$                           & ${>}20$                  \\
      \multicolumn{1}{c}{max(CSV($\Pj_1$),CSV($\Pj_2$))}         & ${>}0.40$                         & ${>}0.50$ (${>}0.244$)     \\
      \multicolumn{1}{c}{min(CSV($\Pj_1$),CSV($\Pj_2$))  }        & ${>}0.40$                         & ${>}0.244$               \\
      \pt(\PH) [\GeVns]                     & ${>}100$                          & ---                     \\
      m(\PH) [\GeVns]                       & ${<}250$                          & $40-250$ (${<}250$)    \\
      m(\PV) [\GeVns]               & ---                              & $75-105$             \\
      \pt(\PV) [\GeVns]                         & $130-180$ (${>}180$)              & $50-100$ (${>}100$)    \\

      \MET [\GeVns]                         & ${>}45$                           & ---                     \\
      \multicolumn{1}{c}{$\Delta\Phi(\MET,\ell)$}      & ${<}\pi/2$                        & ---                     \\
      \hline
    \end{tabular}
\end{table*}

\begin{figure*}[htbp]
  \centering
    \begin{tabular}{cc}
      \includegraphics[width=0.45\textwidth,angle=0]{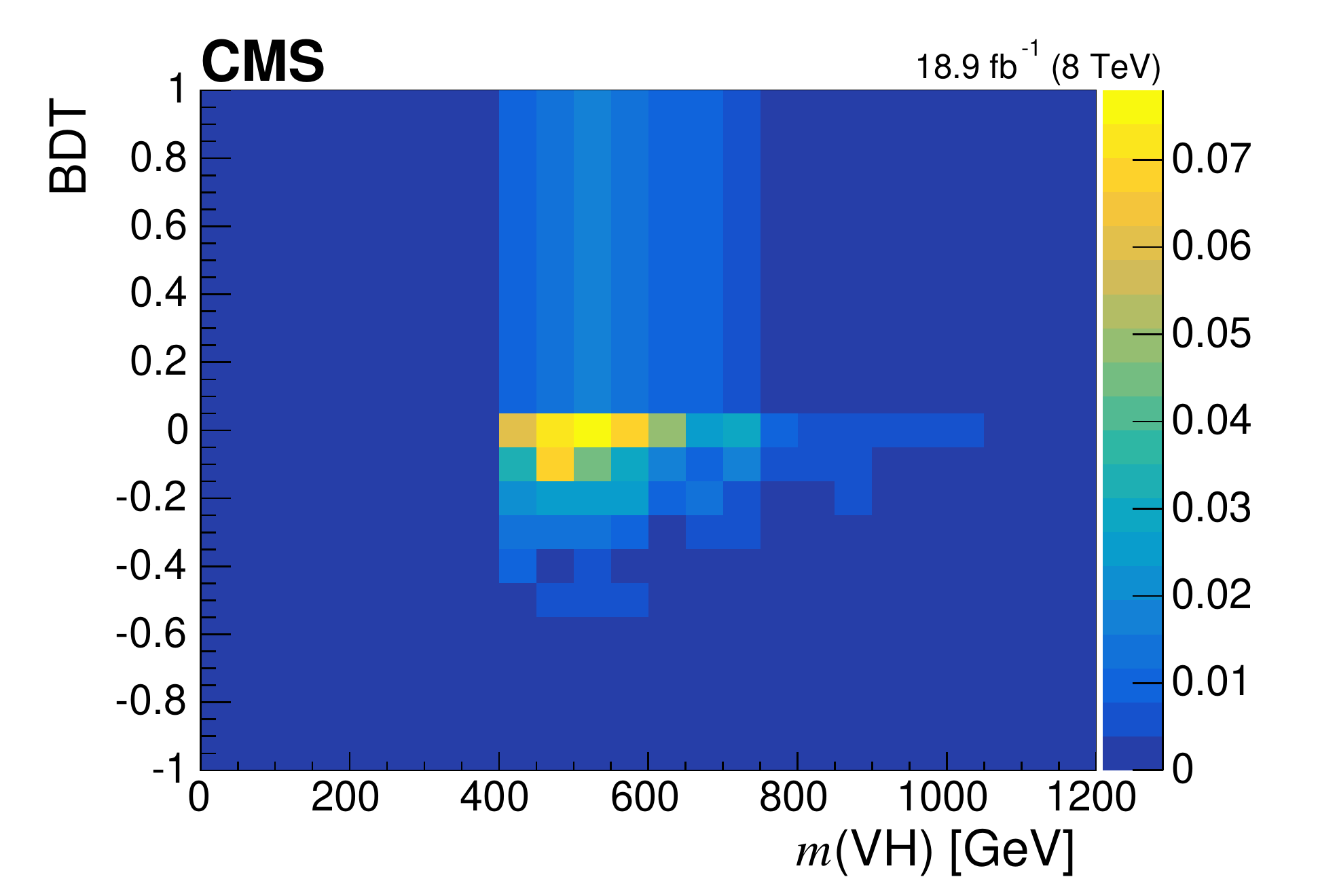} &
      \includegraphics[width=0.45\textwidth,angle=0]{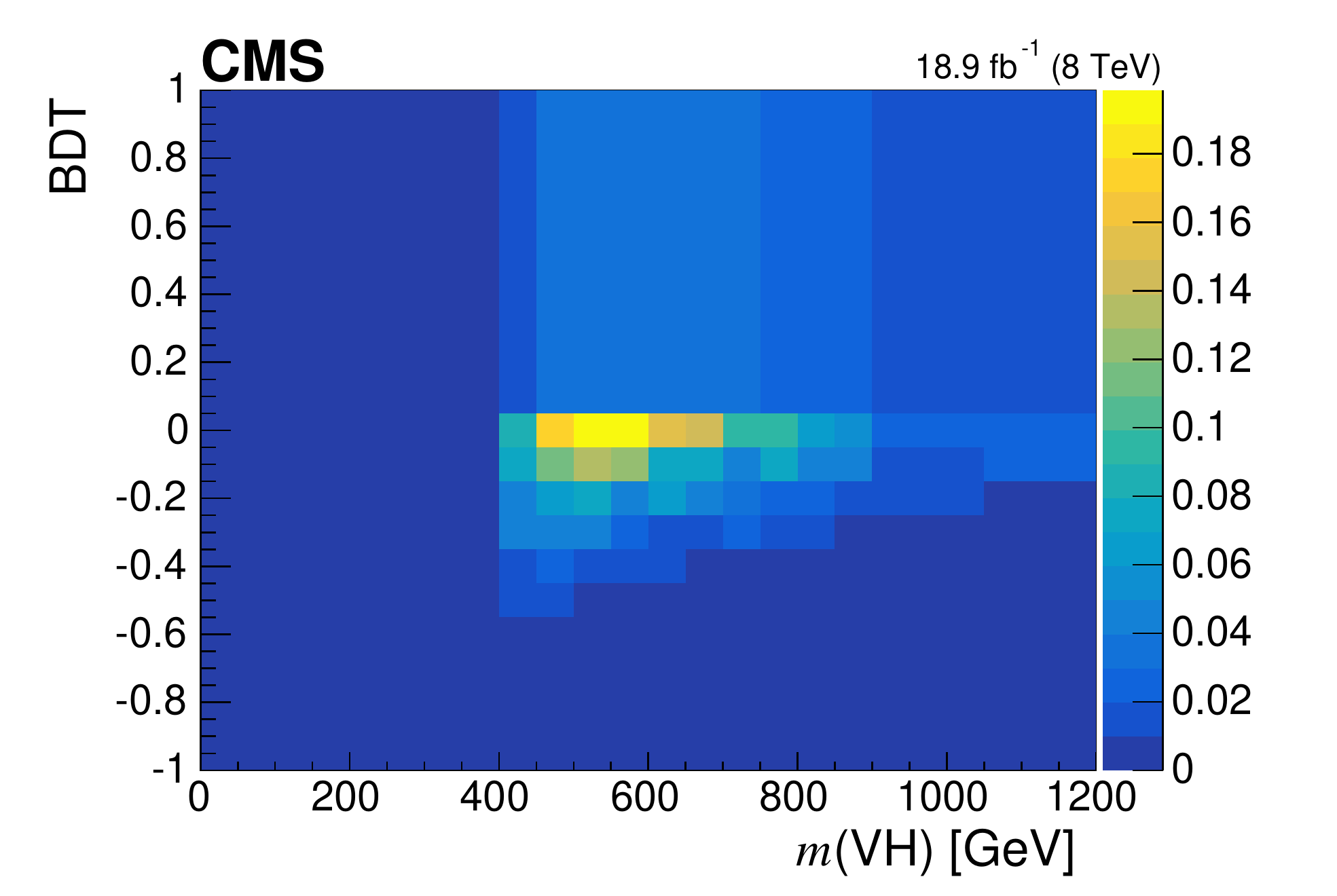} \\
      \multicolumn{2}{c}{\includegraphics[width=0.45\textwidth,angle=0]{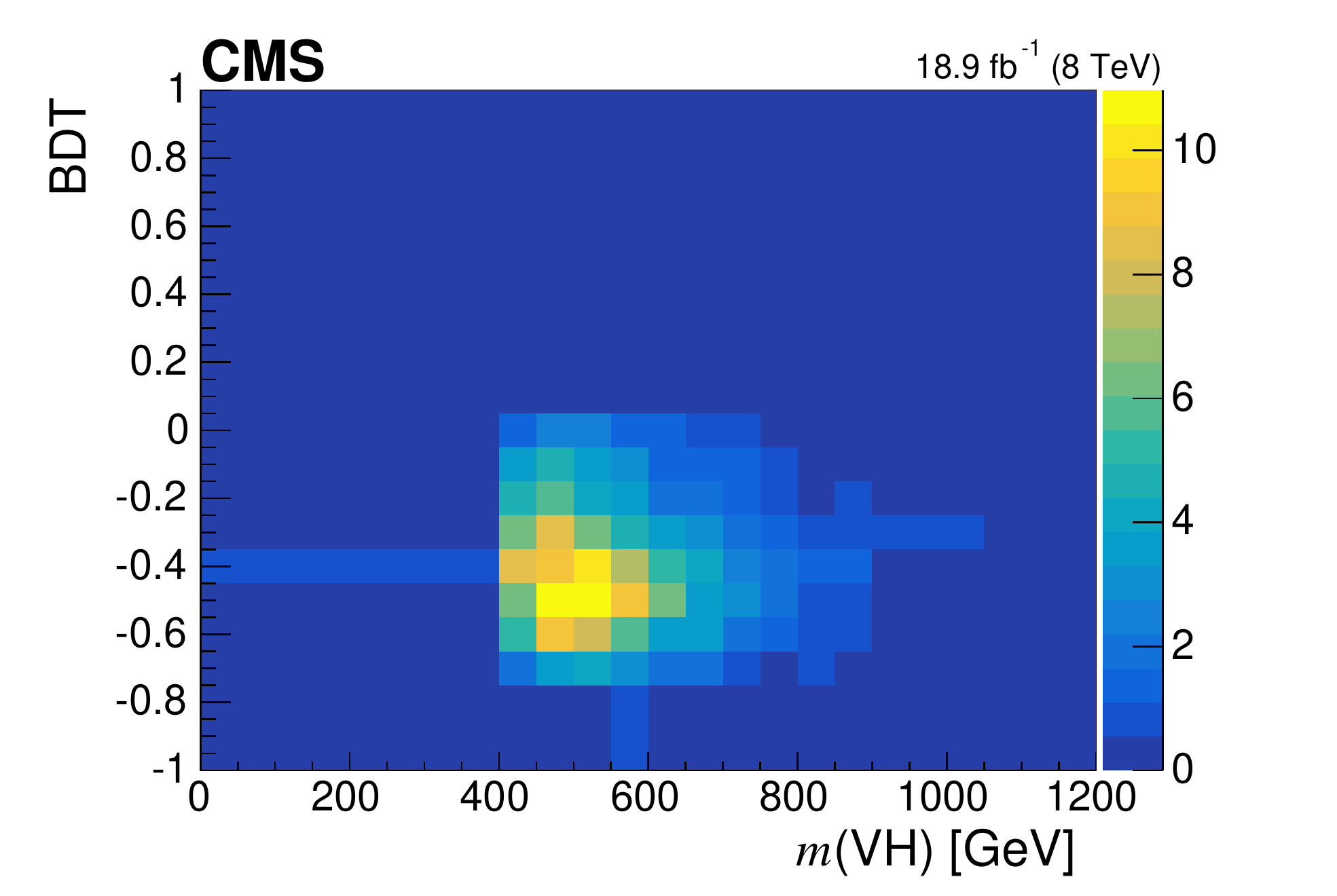}} \\
    \end{tabular}
    \caption{The scalar (left), pseudoscalar (right), and total background (bottom) templates for the high-boost \Wenu channel.  Bin content is normalized according to the bin area.}
    \label{fig:2dtemplates}
\end{figure*}

\begin{figure*}[htbp]
  \centering
    \begin{tabular}{cc}
      \includegraphics[width=0.45\textwidth,angle=0]{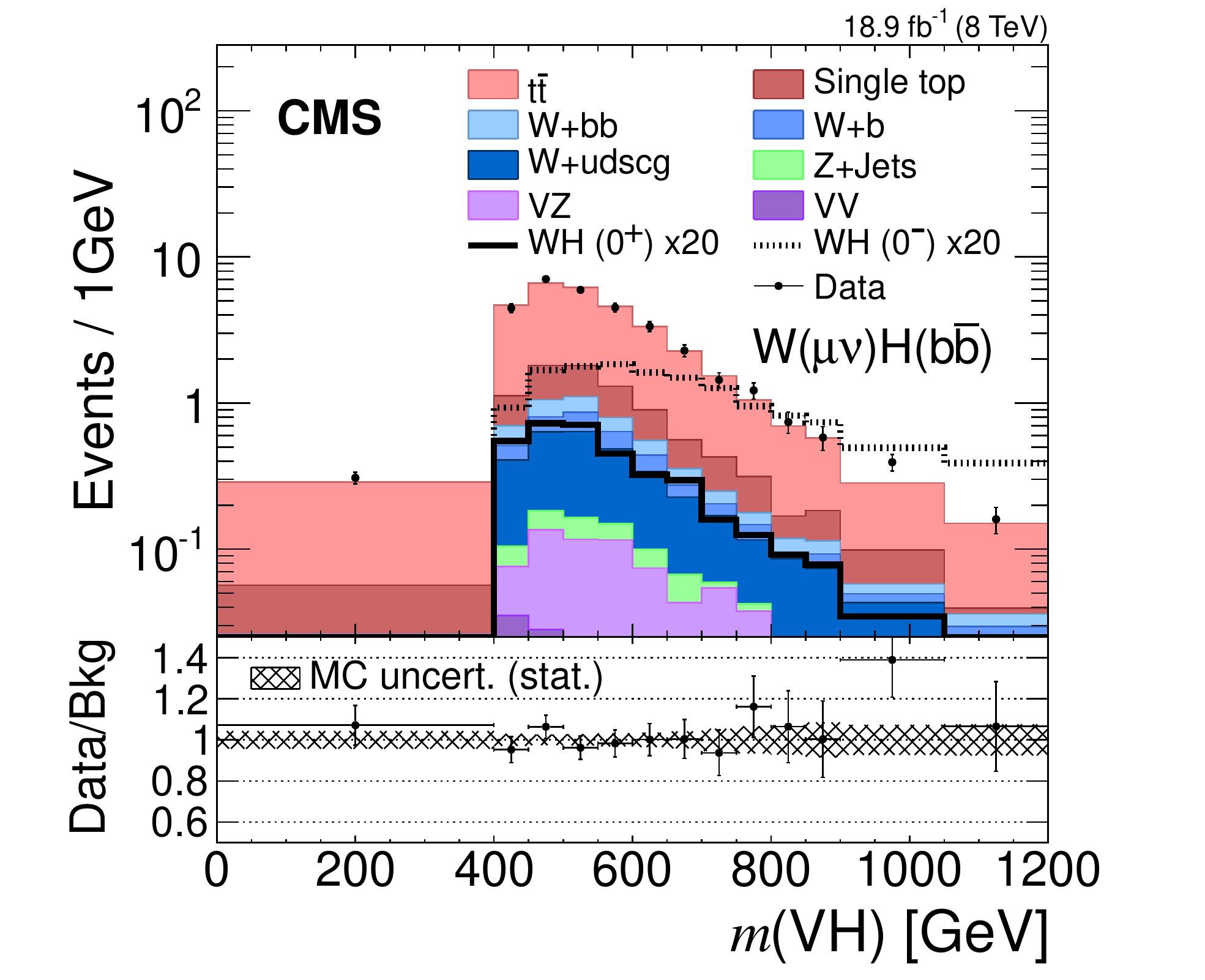} &
      \includegraphics[width=0.45\textwidth,angle=0]{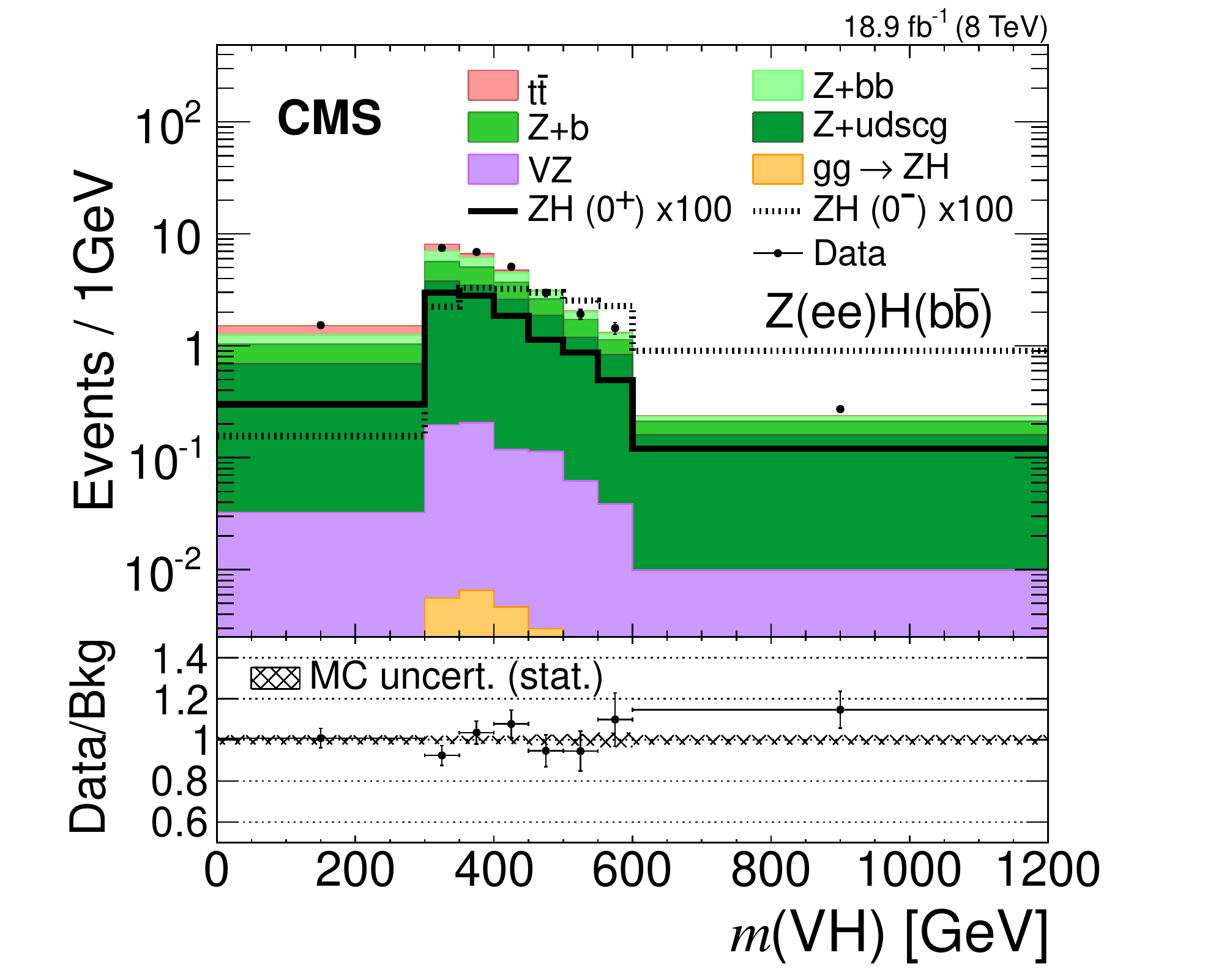} \\
    \end{tabular}
    \caption{The \mVh distributions for the high-boost region of the \Wmunu (left) and \Zee (right) channels.  The distribution observed in data is represented by points with error bars.  SM backgrounds are represented by filled histograms.  A pure scalar (pseudoscalar) Higgs boson signal is represented by the solid (dotted) histogram.  The statistical uncertainty related to the finite size of the simulated background event samples is represented by the hatched region.  Values of $\mVh>1200\GeV$ are included in the last bin.  The bin content is normalized according to the bin width.  The lower panel shows the ratio of the observed and expected background yields.}
    \label{fig:control}
\end{figure*}

The expected scalar, pseudoscalar, and total background templates for the high-boost \Wenu channel are shown in Fig.~\ref{fig:2dtemplates}.  One-dimensional projections of the templates for the high-boost \Wmunu and \Zee channels onto the \mVH axis are shown in Fig.~\ref{fig:control}.  The discrimination power of \mVh for the scalar and pseudoscalar hypotheses can be seen clearly; the pseudoscalar hypothesis tends to produce larger values of \mVh than the scalar hypothesis.

\section{Systematic uncertainties}
\label{sec:systematics}

A variety of sources of uncertainty are considered in this analysis.  These include the energy scale, energy resolution, and reconstruction efficiencies of the relevant physics objects; integrated luminosity determination; cross section and background normalization scale factor uncertainties; and the accuracy and finite size of the simulated event samples.  The treatment of most uncertainties is identical to that of Ref.~\cite{Chatrchyan:2013zna}, with the exceptions discussed below.  All uncertainties are summarized in Table~\ref{tab:systematics}.

Uncertainties are assigned to both the scalar and pseudoscalar signal yields, related to the calculation of higher-order QCD and EW corrections.  In the pseudoscalar case, the uncertainty in the NLO EW corrections is taken to be the size of the corrections for a scalar Higgs boson.  A slight mismodeling of the \mVH distribution is observed in a sideband of the medium-boost regions with values of the BDT discriminant less than $-0.3$.  This sideband has negligible signal content.  The ratio of data to the background prediction has an approximately constant, positive slope.  As a result, an additional \mVh modeling systematic uncertainty is included, which allows for a linear correction of the background model.  The size of this uncertainty is taken as
twice the ratio of data to prediction, as fitted by a linear function in \mVh.

\begin{table*}[tbp]
  \topcaption{Summary of the sources of systematic uncertainty on the background and signal yields.  The size of the uncertainties that only affect normalizations are given.  Uncertainties that also affect the shapes are implemented with template morphing, a smooth vertical interpolation between the nominal shape and systematic shape variations.}
  \label{tab:systematics}
  \centering
    \begin{tabular}{lc} \hline
      Source                                        & Pre-fit uncertainty     \\  \hline\hline
      Normalization uncertainties & \\\hline
      Integrated luminosity                                    & 2.6\%                   \\
      Lepton reconstruction and trigger efficiency  & 3\% per $\ell$    \\
      Missing transverse energy scale and resolution                     & 3\%                     \\
      Signal and background cross section (scale)   & 4--6\%                 \\
      Signal and background parton distribution functions                    & 1\%                     \\
      \zerop (\zerom) EW/QCD signal corrections    & 2\%/5\% (10\%/5\%)      \\
      \ttbar and \PV+jets data-driven scale factors                  	   & 10\%                    \\
      Single top quark cross section                      & 15\%                    \\
      Diboson cross section                         & 15\%                    \\
      $\Pg\Pg\to\PZ\PH$ cross section                            & $^{+35\%}_{-25\%}$      \\\hline \hline
     Normalization + shape uncertainties & \\\hline
      Jet energy scale                              & ${\pm} 1\sigma$                   \\
      Jet energy resolution                         & ${\pm} 1\sigma$                   \\
      b tagging efficiency                          & ${\pm} 1\sigma$                   \\
      b tagging mistag rate                         & ${\pm} 1\sigma$                   \\
      Simulated event statistics                    & ${\pm} 1\sigma$                   \\
      Event simulation accuracy (V+jets and \ttbar )& Alternate event simulation          \\
      \mVh modeling                                 & ${\pm}2\, \times$ fitted slope       \\
      \hline
    \end{tabular}
\end{table*}

\section{Results}
\label{sec:results}

\begin{figure}[htbp]
  \centering
     \includegraphics[width=0.45\textwidth,angle=0]{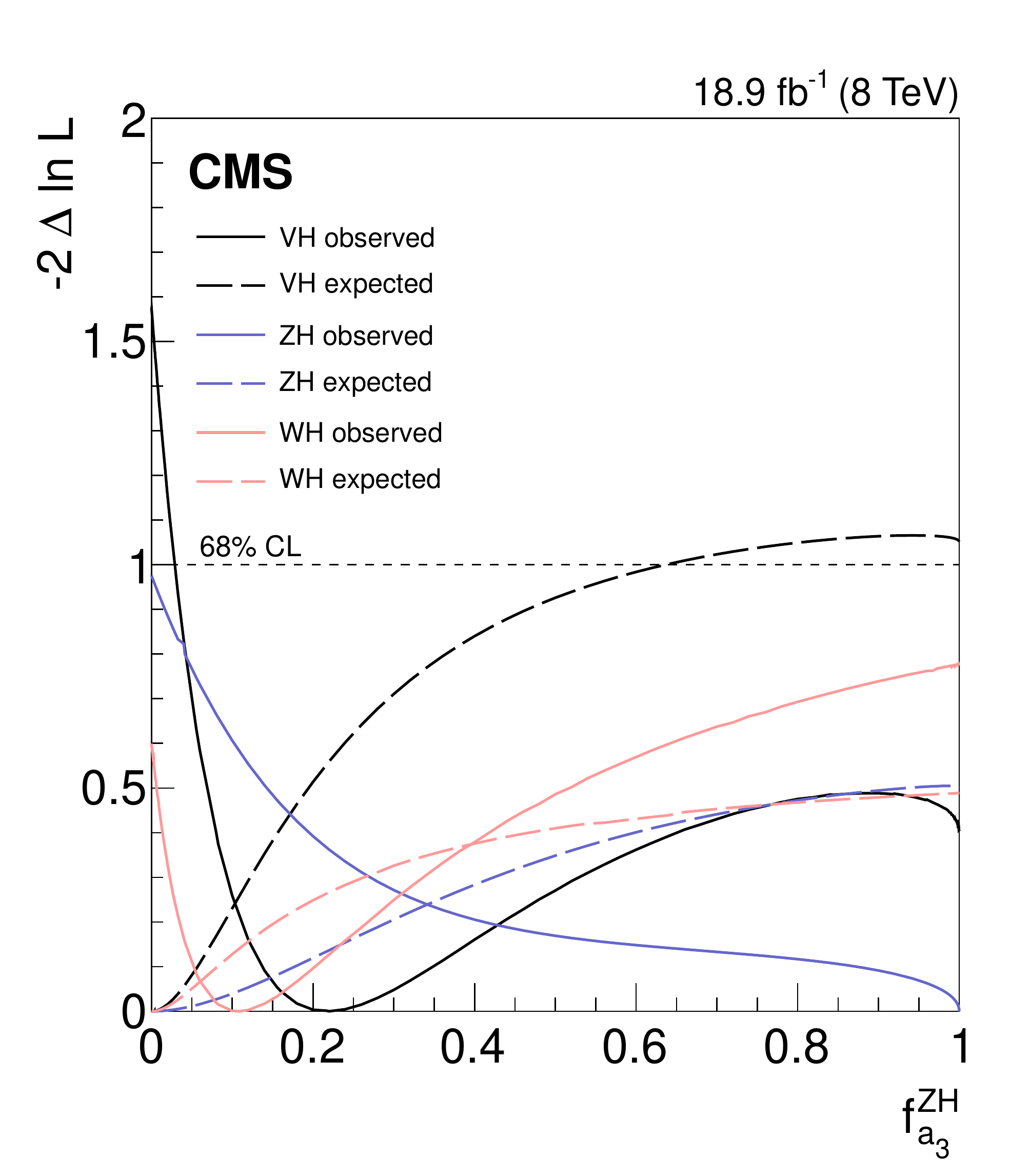}
 	 \caption{Results of profile likelihood scans for the WH and ZH channels, as well as the combination (VH).  The dotted (solid) lines  show the expected (observed) -2$\Delta\mathrm{ln}\mathcal{L}$ value as a function of $\faThreeZH$.  A horizontal dashed line is shown, representing the 68\% CL.}
    \label{fig:VH}
\end{figure}

Results of one-dimensional profile likelihood scans in the VH channels are shown in Fig.~\ref{fig:VH}, in terms of \faThreeZH.  Throughout the paper, expected results are derived from an Asimov data set~\cite{asimov} for a pure scalar Higgs boson with $\mu=1$. This dataset represents the expectation for an SM Higgs boson in the asymptotic limit of large statistics.  The combined VH scan assumes $a_i^{\PH\PW\PW}=a_i^{\PH\PZ\PZ}$.

The expected $-2\Delta\ln\mathcal{L}$ values reach a plateau above $\faThreeZH\approx 0.3$, as a result of the small $\sigma_1/\sigma_3$ values in the VH channels.  Even for modest values of \faThreeZH, the total signal cross section, and therefore the \mVh shape, is dominated by the pseudoscalar contribution.  Increasing \faThreeZH further has little impact on the \mVH shape, and therefore the likelihood.

Based on the available data, the VH channels alone do not have sufficient sensitivity to derive any constraint on \faThree at 95\% CL.  Although there is some discrepancy between the expected and observed scans, all observed results are consistent with the SM prediction of $\faThree=0$.  This discrepancy is driven by a modest excess (deficit) at high (low) values of \mVh in a selected number of background-depleted bins in the high-boost $\PZ\to\Pe\Pe$ and $\PW\to\mu\nu$ channels, which is consistent with the SM prediction within statistical and systematic uncertainties.

\begin{figure*}[htbp]
  \centering
     \begin{tabular}{cc}
        \includegraphics[width=0.45\textwidth,angle=0]{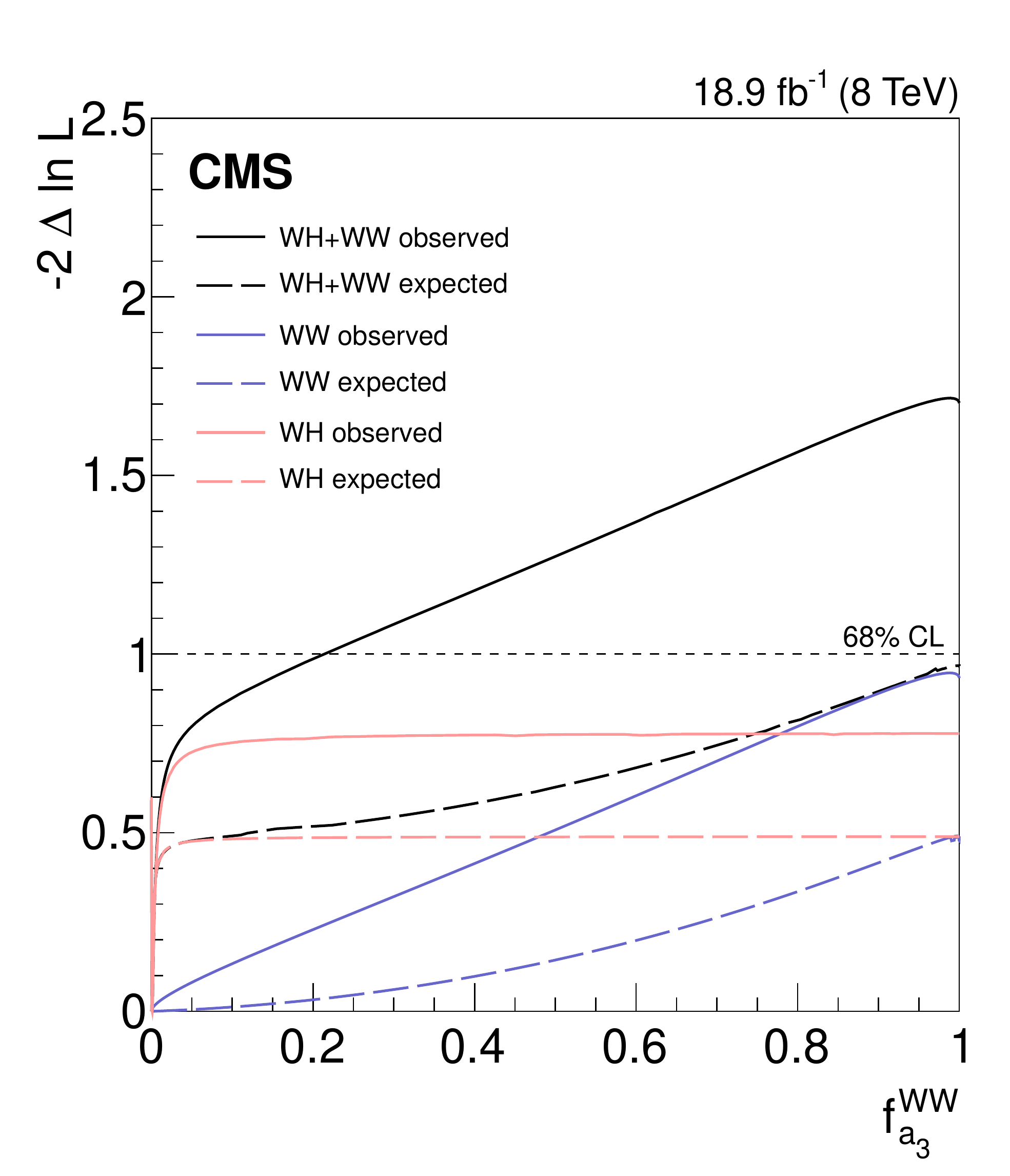} &
        \includegraphics[width=0.45\textwidth,angle=0]{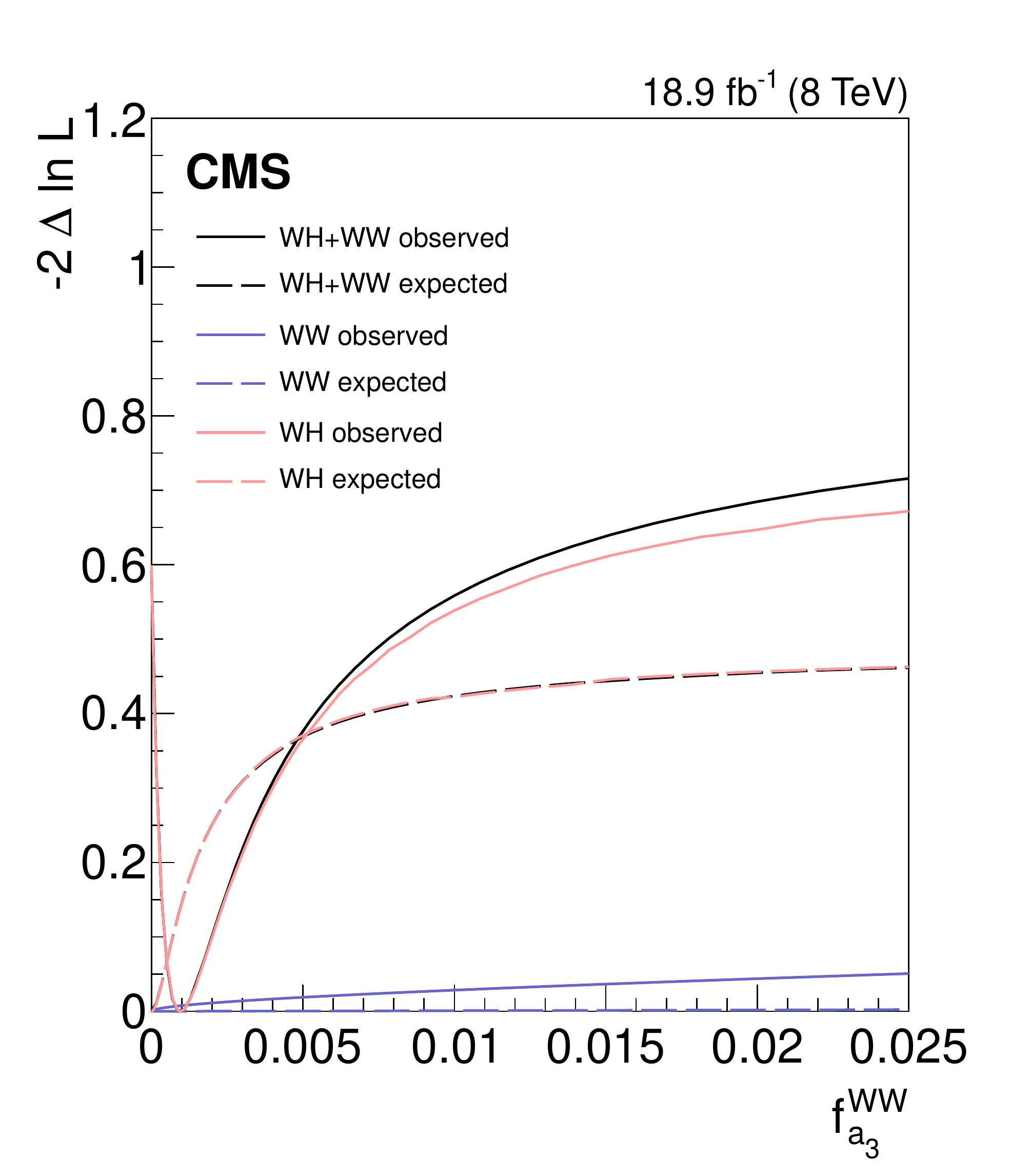} \\
        \includegraphics[width=0.45\textwidth,angle=0]{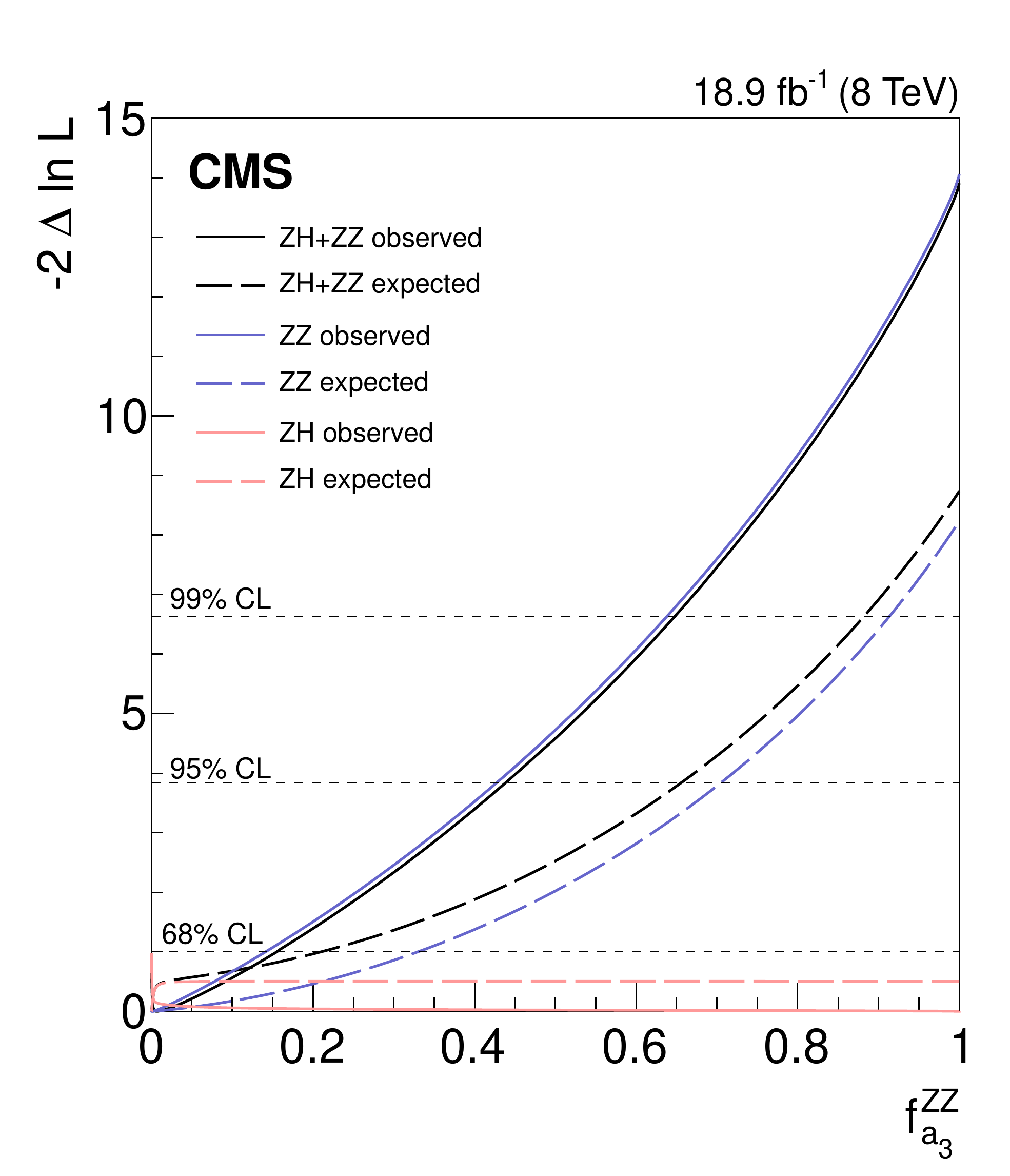} &
		\includegraphics[width=0.45\textwidth,angle=0]{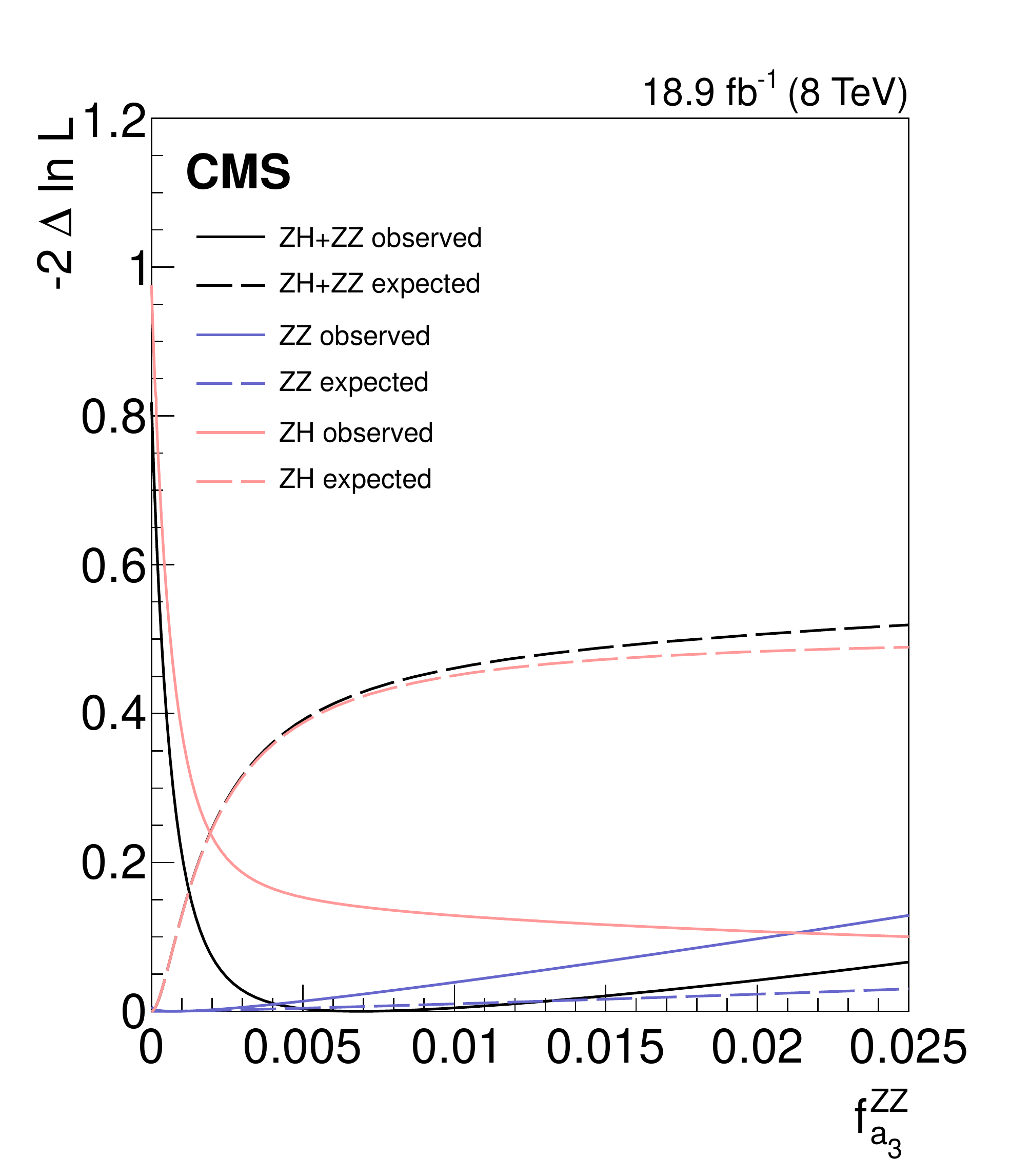}\\
      \end{tabular}
    \caption{Results of profile likelihood scans for the VH and VV channels, plus their combination. The dotted (solid) lines show the expected (observed) -2$\Delta\mathrm{ln}\mathcal{L}$ value as a function of $\faThree$.  The full range of \faThree is shown on the left, with the low \faThree region highlighted on the right. Horizontal dashed lines represent the 68\%, 95\%, and 99\% CL.}
    \label{fig:VVVH}
\end{figure*}

\begin{figure*}[htbp]
  \centering
     \begin{tabular}{cc}
        \includegraphics[width=0.45\textwidth,angle=0]{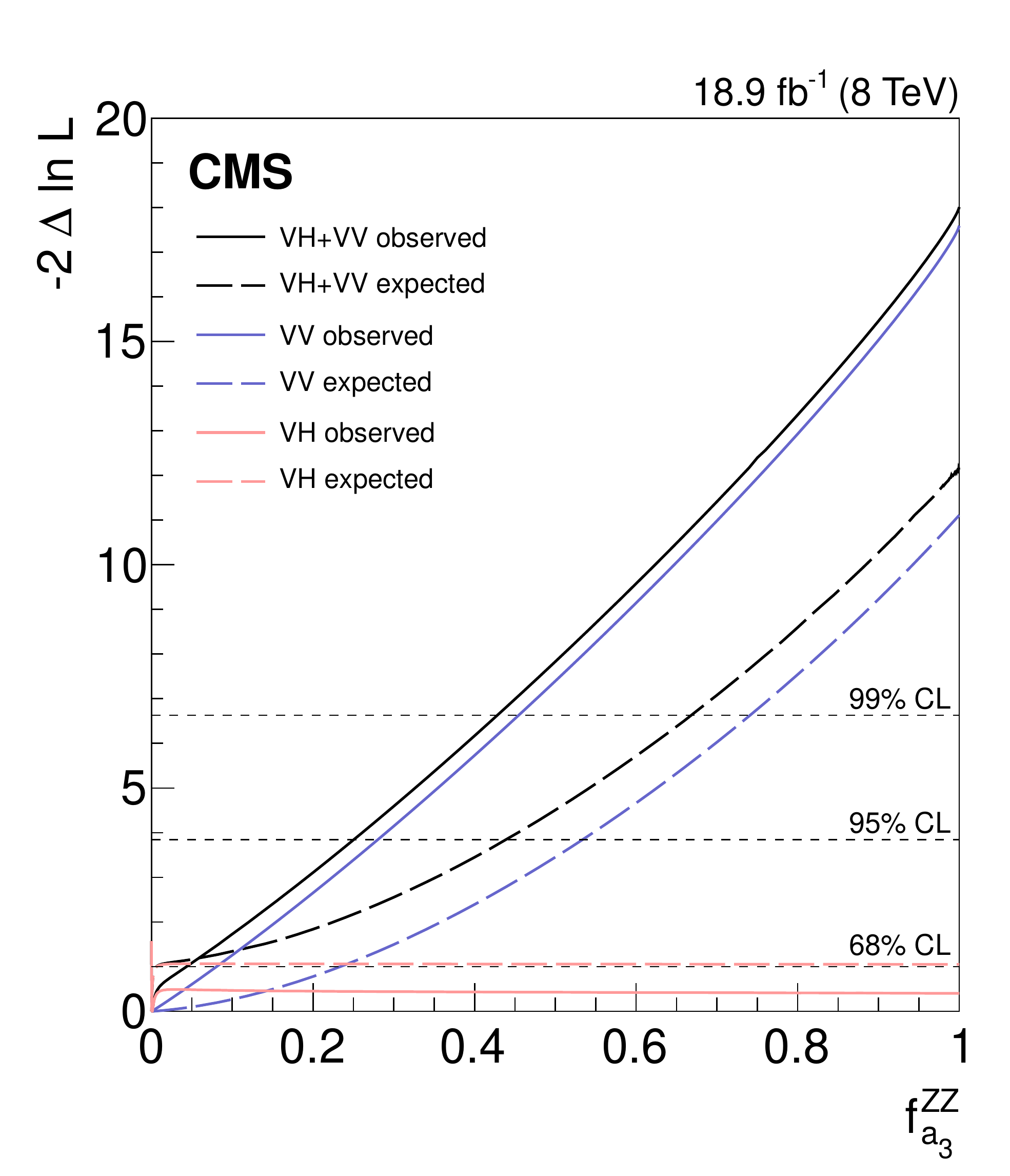} &
        \includegraphics[width=0.45\textwidth,angle=0]{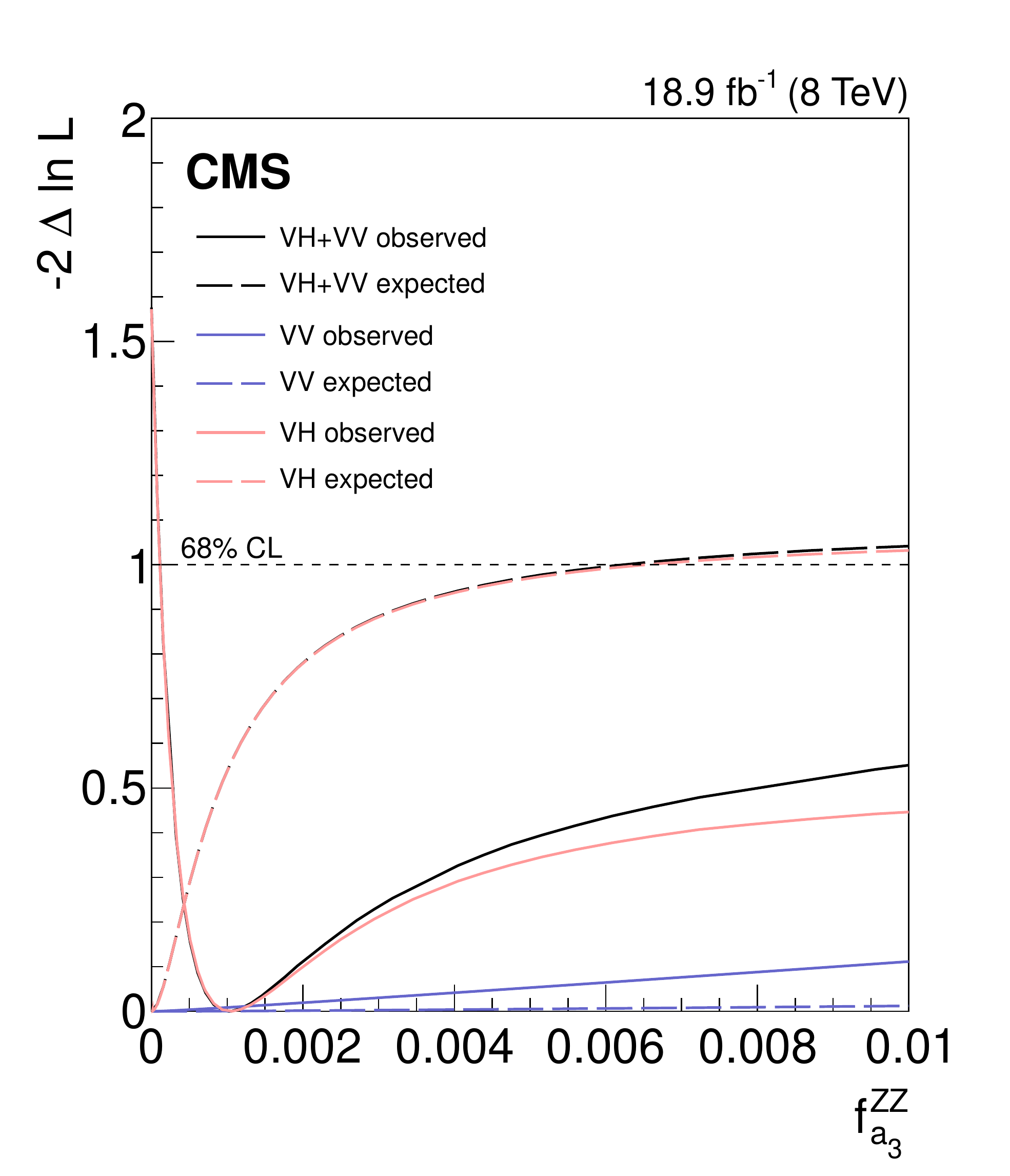} \\
        \includegraphics[width=0.45\textwidth,angle=0]{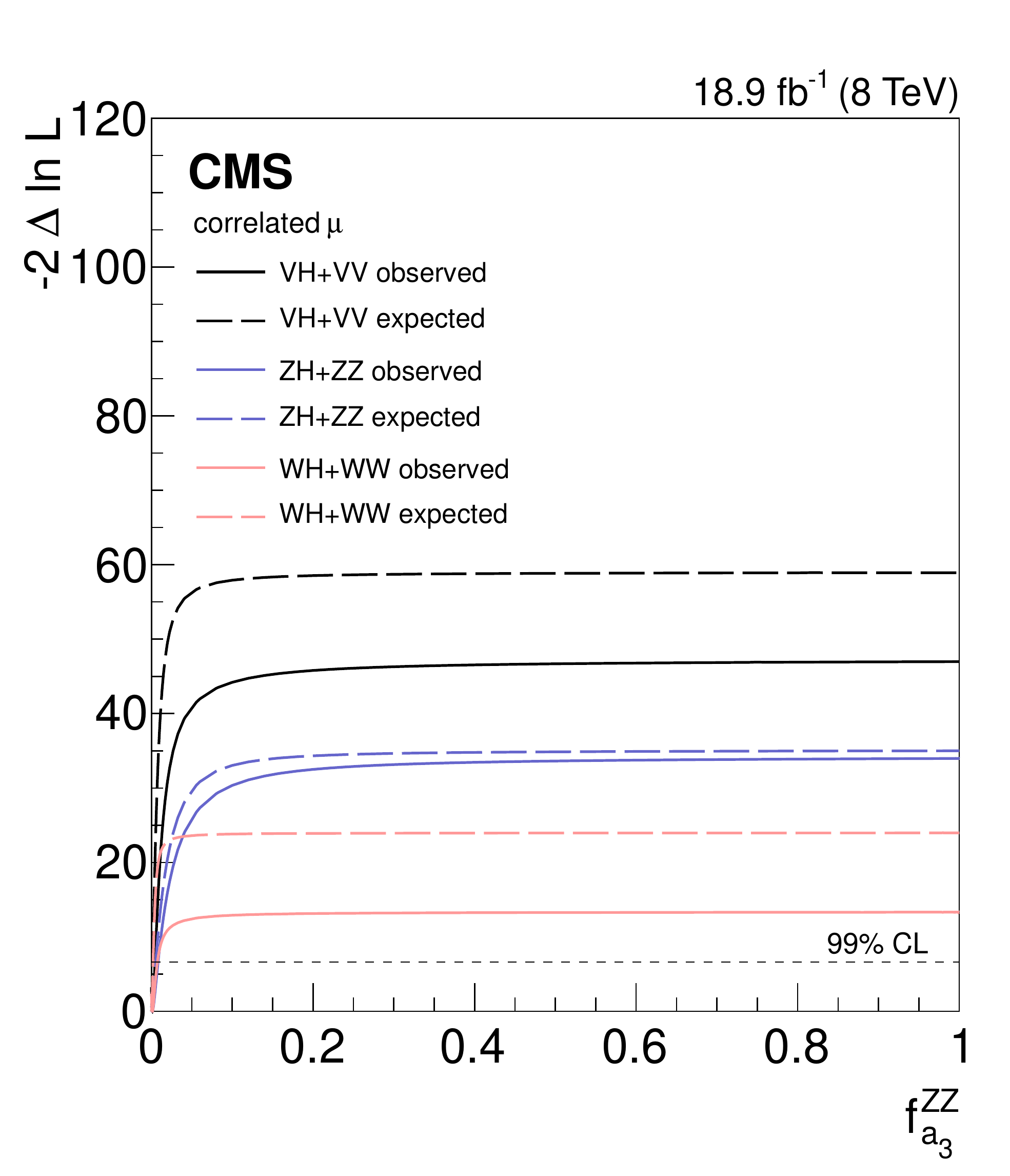} &
        \includegraphics[width=0.45\textwidth,angle=0]{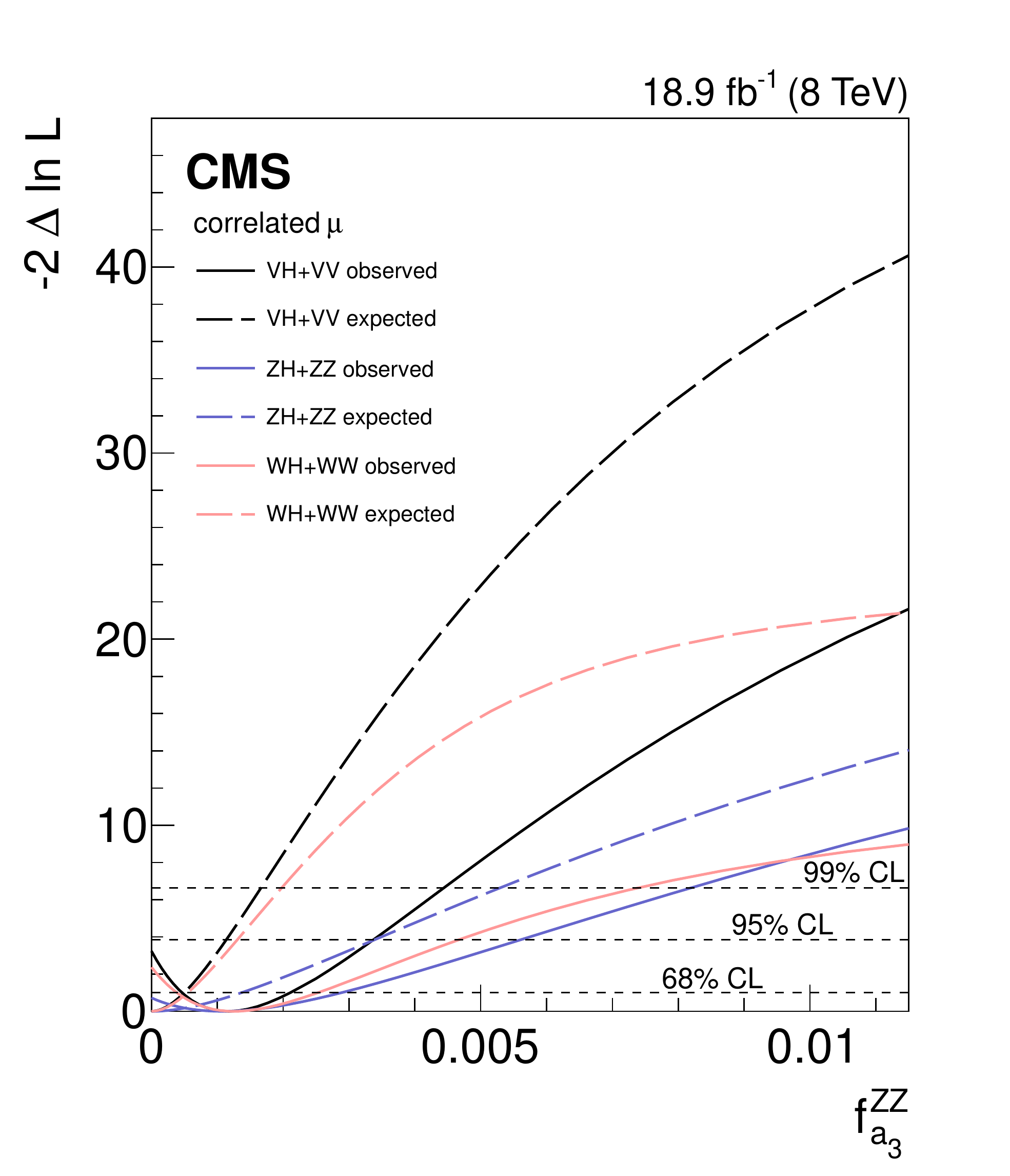}\\
      \end{tabular}
    \caption{Results of profile likelihood scans for the VH and VV channels, as well as their combination. The dotted (solid) lines show the expected (observed) -2$\Delta\mathrm{ln}\mathcal{L}$ value as a function of $\faThree$.  The full range of \faThree is shown on the left, with the low \faThree region highlighted on the right.  The bottom plots contain the results of \cm scans.  Horizontal dashed lines represent the 68\%, 95\%, and 99\% CL.  In the legend, VH refers to the combination of the WH and ZH channels, and VV refers to the combination of the \HtoWW and \HtoZZ channels.}
    \label{fig:VVVH2}
\end{figure*}

Results from the VH channels are combined with results from the \HtoVV channels~\cite{Khachatryan:2014kca},  with and without assuming the SM ratio of the $\PH\PQb\PAQb$ and $\PH\PQt\PAQt$ coupling strengths.
Combined profile likelihood scans are shown in Figs.~\ref{fig:VVVH} and ~\ref{fig:VVVH2}, in terms of \faThreeZZ or \faThreeWW.
The $-2\Delta\ln\mathcal{L}$ distributions shown here for the VH channels alone are the same as those shown in Fig.~\ref{fig:VH}, after a transformation of the $x$-axis to \faThreeWW or \faThreeZZ.  These transformations compress (stretch) the low (high) \faThree region, resulting in the distributions shown.  The position of the $-2\Delta\ln\mathcal{L}$ minima and \faThree confidence intervals are given in Table~\ref{tab:cms}.

The \Wh (\Zh) channel is first combined with the \HtoWW (\HtoZZ) channel, enhancing the sensitivity to anomalous HWW (HZZ) interactions, without the need to introduce any assumption on the relationship between HWW and HZZ couplings.  These results are shown in the upper (lower) portion of Fig.~\ref{fig:VVVH}.
The \HtoWW channel alone is not able to constrain \faThree at 68\% CL.  However, in the \ucm combination of the WH and \HtoWW channels, $\faThreeWW>0.21$ is disfavoured at 68\% CL.  Due to the modest preference in the ZH channel for large \faThree, the \ucm combination of the ZH and \HtoZZ channels results in a bound on \faThree that is slightly weaker than that from the \HtoZZ channel alone.

All four channels are combined under the assumption $a_i^{\PH\PW\PW}=a_i^{\PH\PZ\PZ}$.  The results of this \ucm combination are shown in the top of Fig.~\ref{fig:VVVH2}. A slight improvement over the constraint from the \HtoVV channels alone is observed, with $\faThreeZZ>0.25$ excluded at 95\% CL.

Correlated-$\mu$ combinations of the VH and \HtoVV channels are performed as well, which are based on the assumption of the SM ratio of the $\PH\PQb\PAQb$ and $\PH\PQt\PAQt$ coupling strengths.  This assumption fixes the relationship between the signal strengths in the VH and \HtoVV channels.  As a result of the relatively well measured signal strengths in the \HtoVV channels, for intermediate and large values of \faThree the signal strengths in the VH channels are constrained to large values, and such a signal cannot be accommodated by the data.  The results are shown in the bottom of Fig.~\ref{fig:VVVH2}.  Relative to the \faThree exclusions obtained from the \HtoVV channels alone, the results obtained here are significantly stronger, with $\faThreeZZ>0.0034$ excluded at 95\% CL in the full combination of all channels.

The future power of the VH channels at probing small anomalous $\PH\PV\PV$ couplings is demonstrated on the right side of Figs.~\ref{fig:VVVH} and ~\ref{fig:VVVH2}.  Although the expected exclusion of anomalous couplings in these channels is only at the $\sim$68~\% CL level with the current 8\TeV dataset,
the $-2\Delta\ln\mathcal{L}$ values increase sharply for small, non-zero values of \faThreeZZ and reach a plateau at $\faThreeZZ\approx 0.05$.  With the inclusion of $\sqrt{s}=13\TeV$ collision data from the ongoing LHC run, the shape of these $-2\Delta\ln\mathcal{L}$ distributions will not change significantly, but the plateau will reach larger values of $-2\Delta\ln\mathcal{L}$.  As soon as the exclusion of a pure pseudoscalar becomes possible, it will be possible to exclude small values of \faThreeZZ as well.

\begin{table*}[htb]
  \centering
    \topcaption{A summary of the locations of the minimum -2$\Delta\mathrm{ln}\mathcal{L}$ values in one-dimensional \faThree profile likelihood scans.  Parentheses contain 68\% CL intervals, and brackets contain 95\% CL intervals. The ranges are truncated at the physical boundaries $0<\faThree< 1$.  The results of combinations which involve both VH and \HtoVV channels are given with and without assuming the SM ratio of the coupling strengths of the Higgs boson to top and bottom quarks.}
    \resizebox{\textwidth}{!}{
\renewcommand{\arraystretch}{1.35}
                \label{tab:cms}
            \newcolumntype{a}{D{+}{\ + \ }{5,10}}
              \begin{tabular}{acll} \hline
\multicolumn{1}{c}{Channel}             & Parameter   & Expected  & Observed \\ \hline \hline
\multicolumn{1}{c}{\Vh}                 & \faThreeZH & 0 (0, 0.64) [0, 1] & 0.22 (0.029, 1) [0, 1] \\\hline
\multicolumn{4}{c}{Correlated-$\mu$ combination} \\\hline
\Wh     + \HtoWW    & \faThreeWW & 0 (0, 0.0012) [0, 0.0027] & 0.0026 (0.00082, 0.0053) [0, 0.0098] \\
\Zh     + \HtoZZ    & \faThreeZZ & 0 (0, 0.0014) [0, 0.0034] & 0.0011 (0, 0.0029) [0, 0.0056] \\
\Vh     + \HtoVV    & \faThreeZZ & 0 (0, 0.00050) [0, 0.0011] & 0.0012 (0.00047, 0.0021) [0, 0.0034] \\\hline
\multicolumn{4}{c}{Uncorrelated-$\mu$ combination} \\\hline
\Wh     + \HtoWW    & \faThreeWW & 0 (0, 1) [0, 1] & 0.00088 (0, 0.21) [0, 1] \\
\Zh     + \HtoZZ    & \faThreeZZ & 0 (0, 0.21) [0, 0.66] & 0.0067 (0, 0.16) [0, 0.44] \\
\Vh     + \HtoVV    & \faThreeZZ & 0 (0, 0.0062) [0, 0.44] & 0.0010 (0.00011, 0.043) [0, 0.25] \\\hline
                \end{tabular}
                }
\end{table*}

\begin{figure*}[htbp]
  \centering
    \begin{tabular}{cc}
      \includegraphics[width=0.45\textwidth,angle=0]{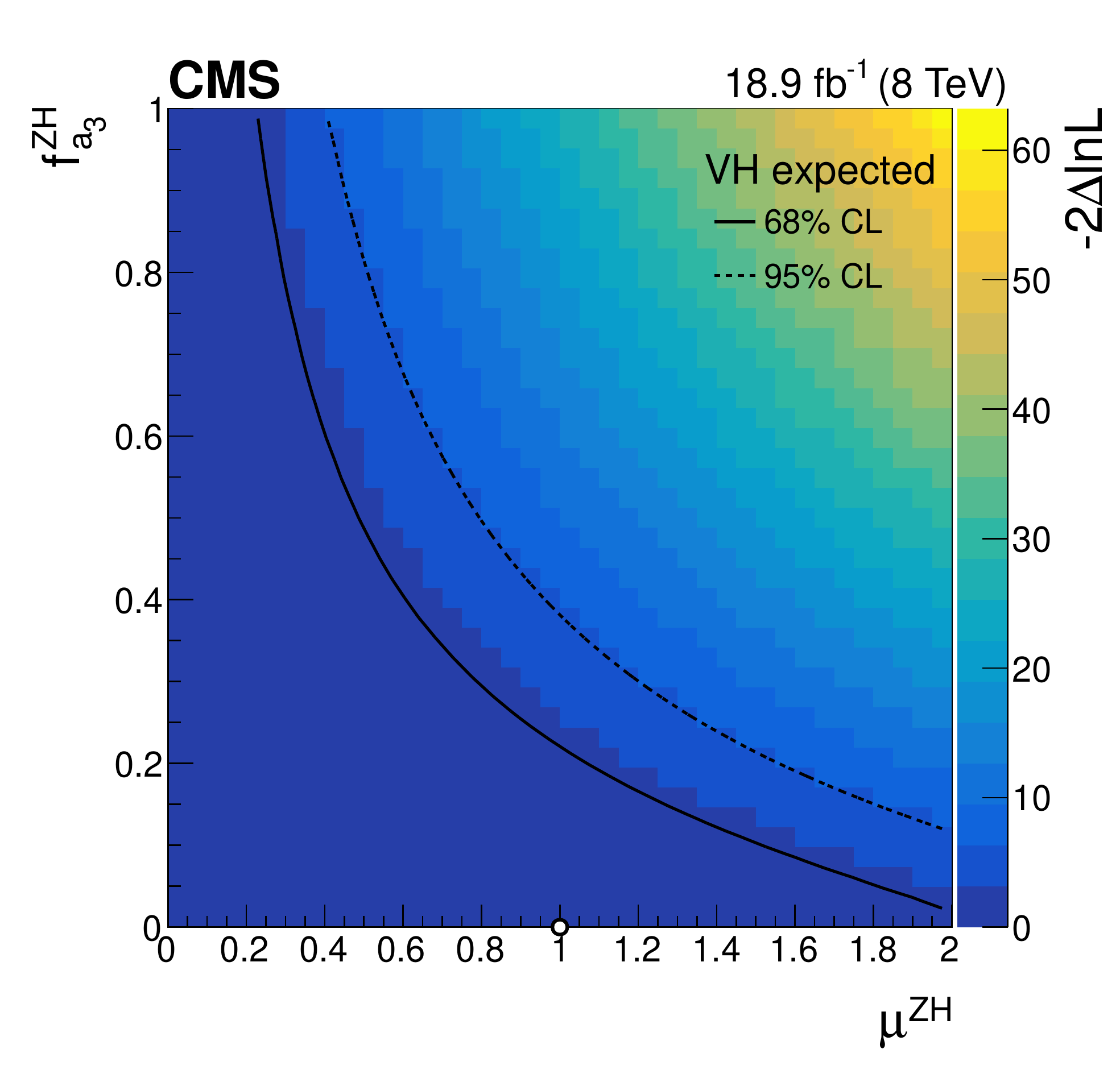} &
       \includegraphics[width=0.45\textwidth,angle=0]{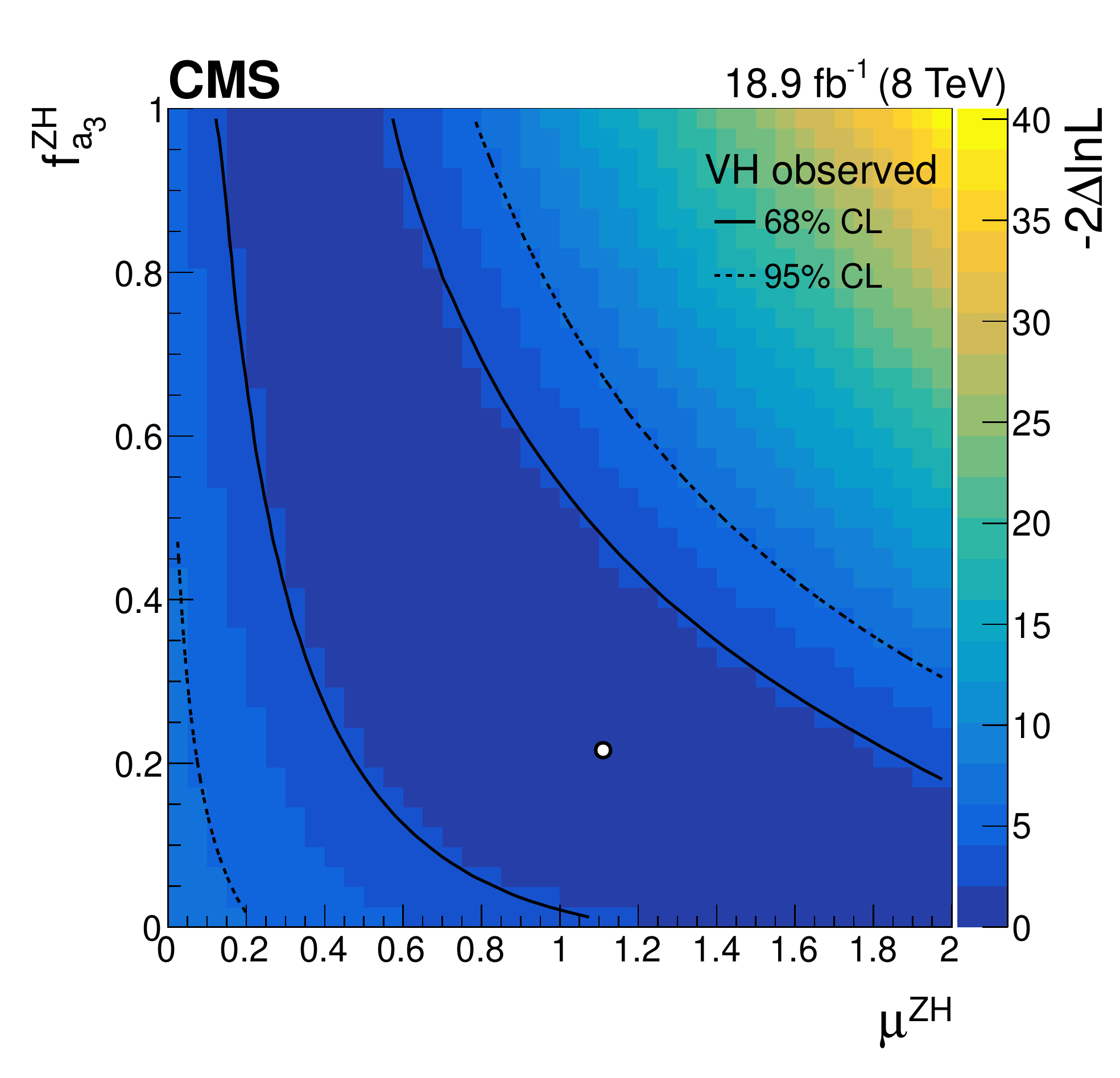} \\
       \end{tabular}
    \caption{Expected (left) and observed (right) two-dimensional profile likelihood scans based on a combination of the WH and ZH channels in the \faThreeZH versus $\mu^{\PZ\PH}$ plane.  The colour coding represents $-2\Delta\mathrm{ln}\mathcal{L}$ calculated with respect to the global minimum.  The scan minimum is indicated by a white dot.  The 68\% and 95\% CL contours at $-2\Delta\mathrm{ln}\mathcal{L}=2.30$ and $5.99$, respectively, are shown.  The observed result includes upper and lower bounds while the expected result contains only upper bounds, as the expected result is consistent with $\faThreeZH=0$ at 68\% CL.}
    \label{fig:2d}
\end{figure*}

Results of two-dimensional profile likelihood scans in the $\mu^{\PZ\PH}$ versus \faThreeZH plane based on a combination of WH and ZH channels are shown in Fig.~\ref{fig:2d}.  Smaller $\mu^{\PZ\PH}$ values are preferred with increasing \faThreeZH as a result of increasing signal efficiency, due to the harder \mVh distribution of a potential pseudoscalar signal compared to that of a scalar.  The minimum of the -2$\Delta\mathrm{ln}\mathcal{L}$ values corresponds to $\mu^{\PZ\PH}=1.11$ and $\faThreeZH=0.22$.

\begin{figure*}[htbp]
  \centering
  \begin{tabular}{cc}
        \includegraphics[width=0.45\textwidth,angle=0]{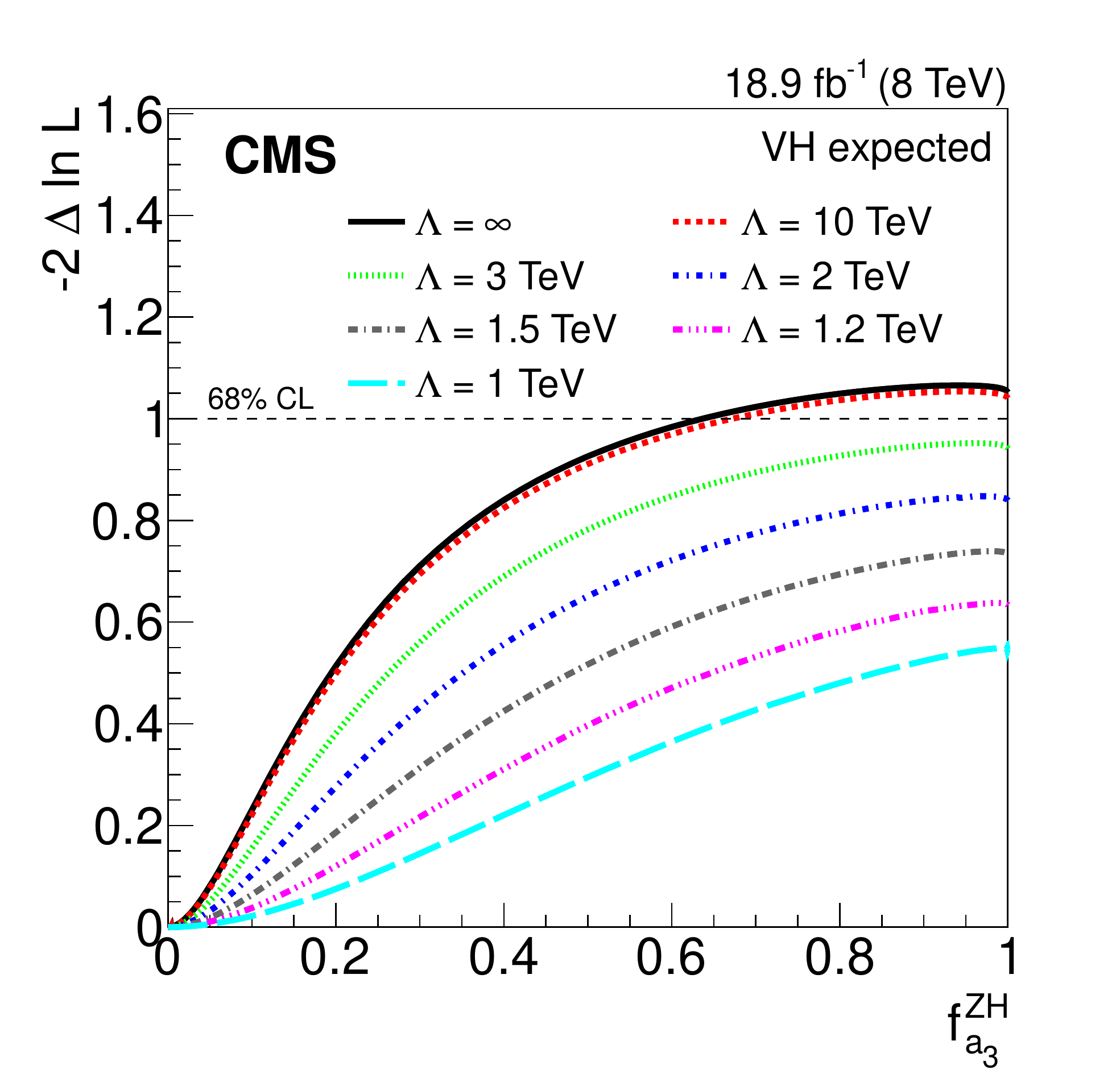} &
        \includegraphics[width=0.45\textwidth,angle=0]{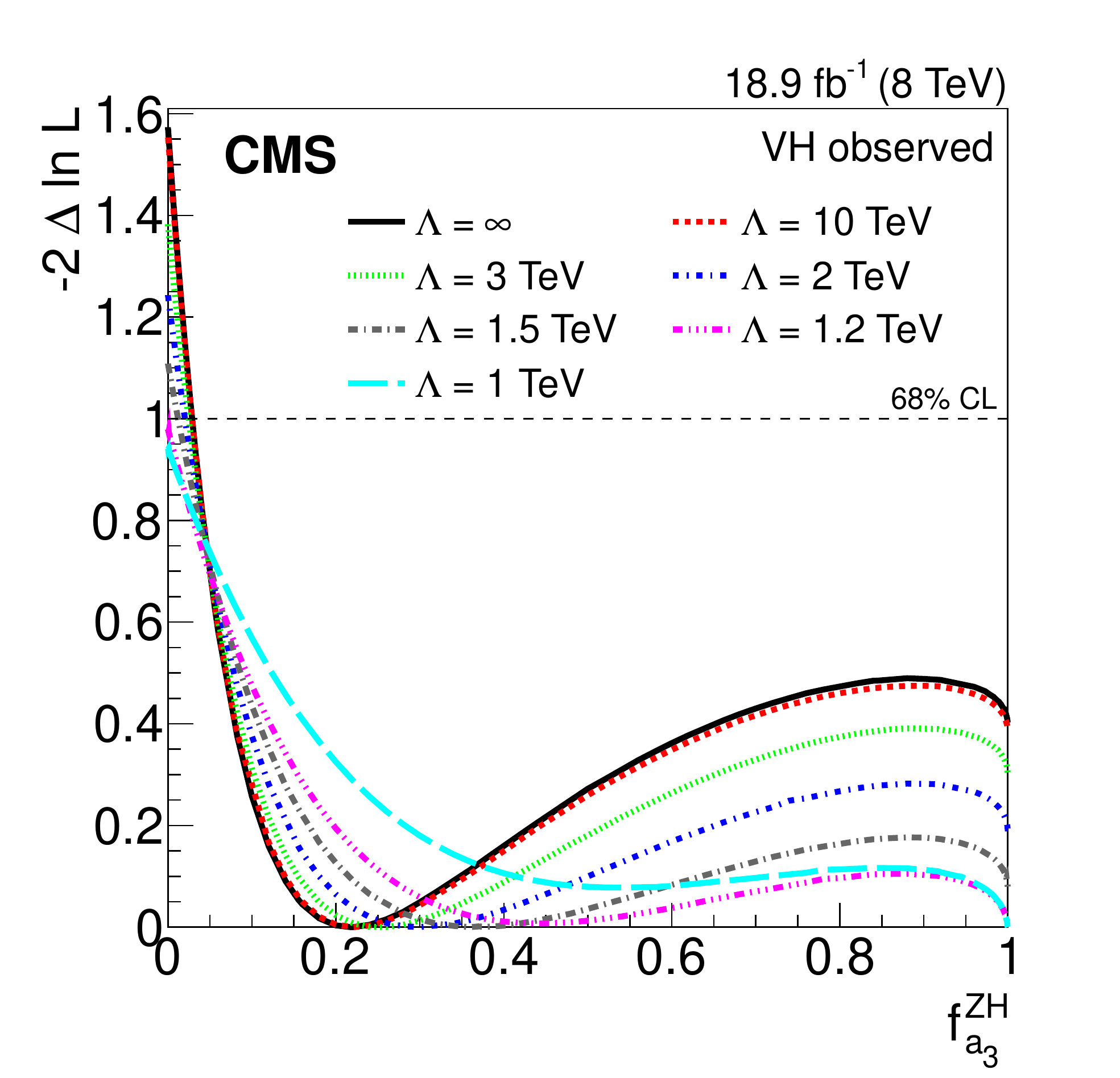} \\
  \end{tabular}
    \caption{Results of expected (left) and observed (right) \faThreeZH scans based on a combination of the WH and ZH channels, with various scales of new physics $\Lambda$.  The coloured lines show the -2$\Delta\mathrm{ln}\mathcal{L}$ value as a function of \faThreeZH.  The horizontal dashed line represents the 68\% CL.}
    \label{fig:lambda}
\end{figure*}

Finally, we allow for the modification of the $a_3^{\PH\PV\PV}$ couplings by a momentum-dependent form factor~\cite{Anderson:2013afp}, given by
\begin{equation}
\left[\left(1+\frac{q_{\PV_1}^2}{\Lambda^2}\right)^2 \, \left(1+\frac{q_{\PV_2}^2}{\Lambda^2}\right)^2\right]^{-1},
\end{equation}
where $\Lambda$ represents a scale of new physics at which the $a_3^{\PH\PV\PV}$ coupling can no longer be treated as a constant.  Unlike earlier results in \HtoVV~\cite{Khachatryan:2014kca} where the vector boson $q^2$ is restricted to $\lesssim 100\GeV$, in VH production much larger values are accessible.  This fact is responsible for much of the sensitivity of this analysis, but also necessitates the consideration of form factor effects.  Profile likelihood scans based on a combination of the WH and ZH channels for various values of $\Lambda$ are shown in Fig.~\ref{fig:lambda}.

For $\Lambda\gtrsim 10$ \TeV, a potential momentum-dependent form factor has a negligible impact on the analysis.  But for smaller values of $\Lambda$, the tail of the \mVh distribution is diminished, and along with it the sensitivity to anomalous couplings.  However, even for $\Lambda$ values as small as 1\TeV, the VH channels maintain significant sensitivity.

\section{Summary}

A search has been performed for anomalous pseudoscalar $\PH\PV\PV$ interactions in $\sqrt{s}=8\TeV$ pp data collected with the CMS detector.  This is the first study of such interactions at the LHC in associated VH production.  The results based on the VH channels are combined statistically with those from a previously published study of \HtoVV decays, which assumes the signal yield is dominated by gluon fusion production of the Higgs boson.  Channels sensitive to the $\PH\PW\PW$ and $\PH\PZ\PZ$ interaction are combined assuming equality of the couplings of the Higgs boson to $\PW$ and $\PZ$ bosons.

A leading order scalar $a_1^{\PH\PV\PV}$ and pseudoscalar $a_3^{\PH\PV\PV}$ coupling with a relative phase of 0 are considered, while all other potential tensor structures are neglected.  The $a_1^{\PH\PV\PV}$ and $a_3^{\PH\PV\PV}$ couplings are first treated as constants, but later modified to allow potential momentum-dependent form factor effects in VH production.  Profile-likelihood scans are used to assess the consistency of the data with various effective pseudoscalar cross section fractions, $\faThree$.

The VH channels alone do not currently have sufficient sensitivity to constrain the $\faThree$ at 95\% CL.  However, $\faThree^{\PZ\PZ}$ can be constrained to the sub-percent level in a combination of VH and \HtoVV channels, when assuming the standard model ratio of the coupling strengths of the Higgs boson to top and bottom quarks.  Under this assumption, and ignoring form factor effects, $\faThree^{\PZ\PZ}>0.0034$ is excluded at 95\% CL in the combination of all channels.

\begin{acknowledgments}

We would like to thank Christoph Englert, Matthew McCullough, and Michael Spannowsky for providing calculations of $\Pg\Pg\to\PZ\PH$ kinematics with non-SM couplings. We especially thank Christoph for his help in understanding the symmetry considerations at work in this process.

We congratulate our colleagues in the CERN accelerator departments for the excellent performance of the LHC and thank the technical and administrative staffs at CERN and at other CMS institutes for their contributions to the success of the CMS effort. In addition, we gratefully acknowledge the computing centres and personnel of the Worldwide LHC Computing Grid for delivering so effectively the computing infrastructure essential to our analyses. Finally, we acknowledge the enduring support for the construction and operation of the LHC and the CMS detector provided by the following funding agencies: BMWFW and FWF (Austria); FNRS and FWO (Belgium); CNPq, CAPES, FAPERJ, and FAPESP (Brazil); MES (Bulgaria); CERN; CAS, MoST, and NSFC (China); COLCIENCIAS (Colombia); MSES and CSF (Croatia); RPF (Cyprus); MoER, ERC IUT and ERDF (Estonia); Academy of Finland, MEC, and HIP (Finland); CEA and CNRS/IN2P3 (France); BMBF, DFG, and HGF (Germany); GSRT (Greece); OTKA and NIH (Hungary); DAE and DST (India); IPM (Iran); SFI (Ireland); INFN (Italy); MSIP and NRF (Republic of Korea); LAS (Lithuania); MOE and UM (Malaysia); CINVESTAV, CONACYT, SEP, and UASLP-FAI (Mexico); MBIE (New Zealand); PAEC (Pakistan); MSHE and NSC (Poland); FCT (Portugal); JINR (Dubna); MON, RosAtom, RAS and RFBR (Russia); MESTD (Serbia); SEIDI and CPAN (Spain); Swiss Funding Agencies (Switzerland); MST (Taipei); ThEPCenter, IPST, STAR and NSTDA (Thailand); TUBITAK and TAEK (Turkey); NASU and SFFR (Ukraine); STFC (United Kingdom); DOE and NSF (USA).

Individuals have received support from the Marie-Curie programme and the European Research Council and EPLANET (European Union); the Leventis Foundation; the A. P. Sloan Foundation; the Alexander von Humboldt Foundation; the Belgian Federal Science Policy Office; the Fonds pour la Formation \`a la Recherche dans l'Industrie et dans l'Agriculture (FRIA-Belgium); the Agentschap voor Innovatie door Wetenschap en Technologie (IWT-Belgium); the Ministry of Education, Youth and Sports (MEYS) of the Czech Republic; the Council of Science and Industrial Research, India; the HOMING PLUS programme of the Foundation for Polish Science, cofinanced from European Union, Regional Development Fund; the OPUS programme of the National Science Center (Poland); the Compagnia di San Paolo (Torino); MIUR project 20108T4XTM (Italy); the Thalis and Aristeia programmes cofinanced by EU-ESF and the Greek NSRF; the National Priorities Research Program by Qatar National Research Fund; the Rachadapisek Sompot Fund for Postdoctoral Fellowship, Chulalongkorn University (Thailand); the Chulalongkorn Academic into Its 2nd Century Project Advancement Project (Thailand); and the Welch Foundation, contract C-1845.

\end{acknowledgments}

\bibliography{auto_generated}

\cleardoublepage \appendix\section{The CMS Collaboration \label{app:collab}}\begin{sloppypar}\hyphenpenalty=5000\widowpenalty=500\clubpenalty=5000\textbf{Yerevan Physics Institute,  Yerevan,  Armenia}\\*[0pt]
V.~Khachatryan, A.M.~Sirunyan, A.~Tumasyan
\vskip\cmsinstskip
\textbf{Institut f\"{u}r Hochenergiephysik der OeAW,  Wien,  Austria}\\*[0pt]
W.~Adam, E.~Asilar, T.~Bergauer, J.~Brandstetter, E.~Brondolin, M.~Dragicevic, J.~Er\"{o}, M.~Flechl, M.~Friedl, R.~Fr\"{u}hwirth\cmsAuthorMark{1}, V.M.~Ghete, C.~Hartl, N.~H\"{o}rmann, J.~Hrubec, M.~Jeitler\cmsAuthorMark{1}, V.~Kn\"{u}nz, A.~K\"{o}nig, M.~Krammer\cmsAuthorMark{1}, I.~Kr\"{a}tschmer, D.~Liko, T.~Matsushita, I.~Mikulec, D.~Rabady\cmsAuthorMark{2}, N.~Rad, B.~Rahbaran, H.~Rohringer, J.~Schieck\cmsAuthorMark{1}, R.~Sch\"{o}fbeck, J.~Strauss, W.~Treberer-Treberspurg, W.~Waltenberger, C.-E.~Wulz\cmsAuthorMark{1}
% ...
\vskip\cmsinstskip
% ...
1:~~Also at Vienna University of Technology, Vienna, Austria\\
2:~~Also at CERN, European Organization for Nuclear Research, Geneva, Switzerland\\
% ...

\end{sloppypar}
\end{document}